\documentclass[conference]{IEEEtran}
\IEEEoverridecommandlockouts
\usepackage{amssymb}
\usepackage{hyperref}
\usepackage{cite}
\usepackage{graphicx}
\usepackage{array}
\usepackage{multirow}
\usepackage{amsmath}
\usepackage{stfloats}
\usepackage{graphicx}
\usepackage{subfigure}
\usepackage{tabularx}
\usepackage{epsfig,epsf}
\usepackage{setspace}
\usepackage{bm}
\usepackage{textcomp}
\usepackage{algorithmic}
\usepackage{algorithm}
\usepackage{subfigure}
\usepackage{caption}
\usepackage{graphicx}
\usepackage{setspace}
\usepackage{url}
\usepackage{verbatim}
\usepackage{graphicx}
\usepackage{cite}
\usepackage{amssymb}
\usepackage{color}
\usepackage{xpatch}
\definecolor{shadecolor}{RGB}{0,0,255}
\definecolor{blue}{RGB}{0,0,255}

\newtheorem{theorem}{Theorem}

\newtheorem{corollary}{Corollary}
\newtheorem{remark}{Remark}
\usepackage{soul}
\makeatletter

\ExplSyntaxOn
\cs_new:Npn \bibColoredItems #1#2
{
	\clist_map_inline:nn {#2} { \cs_new:cpn {bib@colored@##1} {#1} } 
}
\ExplSyntaxOff

\newcommand\bib@setcolor[1]{%
	\ifcsname bib@colored@#1\endcsname
	\expanded{\noexpand\color{\csname bib@colored@#1\endcsname}}%
	\else
	\normalcolor
	\fi
}

\makeatother

\begin{document}
	
\title{Rydberg Atomic Quantum Satellites for Enhanced Ground-to-Space Direct Uplink Access}

\author{Qihao Peng, 
        Tierui Gong,~\IEEEmembership{Member,~IEEE},
        Zihang Song,~\IEEEmembership{Member,~IEEE},
         Qu Luo,~\IEEEmembership{Member,~IEEE},\\
        Cunhua Pan,~\IEEEmembership{Senior Member,~IEEE}, 
        Pei Xiao,~\IEEEmembership{Senior Member,~IEEE}, 
        Chau Yuen,~\IEEEmembership{Fellow,~IEEE}.
   
		\thanks{Q. Peng, Q. Luo, and P. Xiao are affiliated with 5G and 6G Innovation Centre, Institute for Communication Systems (ICS) of the University of Surrey, Guildford, GU2 7XH, UK. (e-mail: \{q.peng, q.u.luo,p.xiao\}@surrey.ac.uk). } \\
       \thanks{Z. Song is with the Centre for Intelligent Information Processing Systems (CIIPS), Department of Engineering, King’s College London, London WC2R 2LS, U.K. (e-mail: zihang.song@kcl.ac.uk).}
          \thanks{C. Pan is with the National Mobile Communications Research Laboratory, Southeast University, Nanjing, China. (e-mail: cpan@seu.edu.cn).} \\
        \thanks{T. Gong and C. Yuen are with the School of Electrical and Electronics Engineering, Nanyang Technological University, Singapore 639798 (e-mail: trgTerry1113@gmail.com, chau.yuen@ntu.edu.sg). (\emph{Corresponding Author: Zihang Song and Tierui Gong.}) }}

	
	\maketitle

\begin{abstract}
	This paper investigates the performance advantages of Rydberg atomic quantum (RAQ)-based multiple-input multiple-output (MIMO) satellites for enhancing direct ground-to-space uplink access. We analytically evaluate the impact of Rydberg atoms on channel estimation by deriving closed-form expressions for the mean-square error (MSE) and normalized mean-square error (NMSE). Based on the estimated channels, we further derive lower bounds on the achievable data rates for maximum ratio combining (MRC) and zero-forcing (ZF) detection schemes. Rigorous analysis demonstrates that RAQ–MIMO outperforms conventional radio-frequency (RF) MIMO under both Rayleigh and satellite channel conditions. Specifically, compared with conventional MIMO, RAQR achieves a ``squaring" gain under Rayleigh fading, especially in long-distance transmission scenarios with stringent power constraints. In contrast, under line-of-sight (LoS)-dominated satellite channels, this gain saturates as channel-estimation benefits diminish, with the remaining improvement primarily arising from the normalized noise background. Monte Carlo simulations validate the analytical results and show that the performance gains of RAQ–MIMO satellites translate into smaller antenna apertures, lower transmit power, and longer communication ranges, thereby paving the way for next-generation satellite networks.
\end{abstract}	
    

\section{Introduction}
Satellite communications are emerging as a key paradigm for 6G owing to their ability to provide seamless coverage and ubiquitous wireless connectivity in remote and rural areas \cite{wu2024large,zuo2024integrating}. In particular, low Earth orbit (LEO) satellites have attracted significant attention because of their advantages in end-to-end latency, path loss, and deployment cost compared with medium Earth orbit (MEO) and geostationary Earth orbit (GEO) counterparts. Motivated by these benefits, major industry practitioners such as Starlink \cite{spacex-starlink} and OneWeb \cite{kokkoniemi2024mission} have launched large-scale projects to deploy ultra-dense LEO satellite constellations, aiming to deliver seamless and ubiquitous global broadband service.

To simultaneous serve multiple users over shared time–frequency resources, multiple-input multiple-output (MIMO) techniques have been extensively studied for LEO satellite systems \cite{Li2020JSAC,KeXin2023TWC,xiang2024massive,shen2022random,Ren2025TWC,wang2025afdm,Zhu2025TWC,song2023ris,Seong2025TSP,Peiwen2025JSAC}. At the physical layer, Le \emph{et al.} exploited frequency reuse and proposed a space–angle–based user-grouping scheduler to mitigate inter-user interference \cite{Li2020JSAC}. For channel estimation, an efficient graph-based pilot allocation and a two-stage estimator were introduced in \cite{KeXin2023TWC}. Based on imperfect channel state information (CSI), the uplink performance of multi-satellite MIMO with statistical CSI was analyzed in \cite{xiang2024massive}. Regarding waveform design, the orthogonal time frequency space has been adopted in satellite communications by exploiting channel sparsity in the delay–Doppler–angle domain \cite{shen2022random,Ren2025TWC}, while Wang \emph{et al.} \cite{wang2025afdm} applied affine frequency division multiplexing with a tailored preamble sequence. Beyond signal processing, expanding spatial degrees of freedom via movable arrays \cite{Zhu2025TWC} and reconfigurable intelligent surfaces \cite{song2023ris} enhances beam coverage and suppresses residual interference. For multiple access, a rate-matching framework based on rate-splitting multiple access was proposed to improve inter-satellite fairness \cite{Seong2025TSP}. Finally, at the service layer, semantics-enabled satellite networking leverages foundation-model segmentation and reconstruction to compress task-relevant information, translating physical-layer gains into end-to-end system efficiency \cite{Peiwen2025JSAC}. Despite these advances, direct ground-to-space access for handheld terminals remains a major challenge due to the stringent constraints on link budget, antenna aperture, and transmit power. 


To address this challenge, quantum sensing provides a promising solution for detecting weak radio-frequency (RF) signals by leveraging quantum phenomena to achieve unprecedented precision in measuring physical quantities \cite{hanzo2025quantum,gisin2007quantum,degen2017quantum}. Particularly, Rydberg atoms possess exaggerated atomic properties such as large electric dipole moments and strong polarizability \cite{sedlacek2012microwave}, enabling interaction with external electromagnetic fields across a broad frequency range from direct current (DC) to terahertz (THz) bands. Furthermore, by exploiting quantum effects such as electromagnetically induced transparency (EIT) \cite{finkelstein2023practical} and Autler–Townes splitting (ATS) \cite{hao2018transition}, incident weak RF signals can be detected optically through a photodetector. These remarkable capabilities establish a bridge between quantum physics and classical wireless communication. By enabling RF-to-optical conversion with quantum-limited sensitivity, Rydberg atomic quantum receivers (RAQRs) exhibit ultra-high sensitivity, broadband tunability, narrowband selectivity, and wide spectral coverage.

Building on these appealing characteristics, extensive research has been devoted to exploring the RAQRs for classical wireless communication \cite{gong2024rydberg,gong2025rydberguplink,zhu2025general,zhu2025raqmimomimomultibandrydberg,cui2025towards,song2025csi} and sensing \cite{chen2025polarization,kim2025multi}. Specifically, Gong \emph{et al.} established an equivalent baseband model for single-input single-output (SISO) RAQR and demonstrated considerable performance gain over conventional RF receivers. This model was later extended to a MIMO architecture \cite{gong2025rydberguplink}, revealing significant improvements in spectral efficiency and energy efficiency compared with conventional MIMO systems. To characterize the reception of dynamic signals, a dynamic signal response model of RAQRs was presented in \cite{zhu2025general}. which was subsequently applied to design multi-band RAQ-MIMO systems \cite{zhu2025raqmimomimomultibandrydberg}. In contrast to the equivalent complex channel models in \cite{gong2024rydberg,gong2025rydberguplink,zhu2025general,zhu2025raqmimomimomultibandrydberg}, Cui \emph{et al.} investigated a magnitude-only model and developed corresponding signal detection algorithms\cite{cui2025towards}. Beyond communication, the sensing capabilities of RAQRs have also been demonstrated in polarization sensing \cite{chen2025polarization} and multi-band angle-of-arrival estimation \cite{kim2025multi}. 

Although existing work has demonstrated the superiority of RAQRs compared to conventional RF MIMO systems, a central open question is whether the ultra-high sensitivity can, under realistic link budgets, enable direct ground-to-satellite access. Beyond connectivity, the system-level benefits of this sensitivity remain uncharacterized, including its potential to reduce antenna array size, extend feasible propagation distance, and improve channel-estimation accuracy in the presence of imperfect CSI.

To the best of our knowledge, this is the first work to investigate direct uplink access in RAQ-MIMO satellite systems, aiming to unlock the potential of RAQRs for satellite networks. Specifically, our contributions are summarized as follows:
\begin{itemize}
	\item We first characterize the impact of integrating RAQ-MIMO into satellite communication networks by deriving closed-form expressions of mean square error (MSE) and normalized mean square error (NMSE) for channel estimation. The analysis reveals a \emph{power-independent} enhancement term governed by the RAQR-specific degrees of freedom \(\rho\) and \(\Phi\), where \(\rho\) is the received gain and \(\Phi\) is the phase shift caused by the local oscillator (LO) and user signals. 
	\item Based on the estimated channels, we derive closed-form lower bounds on the uplink achievable rates for both maximum ratio combining (MRC) and zero-forcing (ZF) schemes, and evaluate these bounds under various scenarios. We quantify the performance degradation due to imperfect CSI and observe that ZF benefits more from enhanced CSI accuracy than MRC. Furthermore, we show that the classical MIMO power-scaling law holds for both Rayleigh and Rician channels, with the achievable rate increasing by a factor determined by the normalized noise background, thereby confirming the superiority of RAQ-MIMO over conventional RF MIMO systems.
	\item We analytically quantify the performance gains of RAQ-MIMO over conventional RF MIMO. In the high-power regime, RAQ-MIMO provides a constant gain for ZF detection determined by the normalized-noise ratio, whereas the performance gain for MRC tends to saturate. In the low-power regime, RAQ-MIMO exhibits a ``squaring" gain for Rayleigh channels, which is particularly beneficial for energy-constrained direct ground-to-space access. However, for typical satellite networks, as the channel becomes more deterministic, this effect diminishes and reduces to a constant gain related to the normalized noise background. These gains also translate into tangible link-budget advantages, including reductions in transmit power and antenna apertures, as well as significantly extended coverage.

	\item Monte Carlo simulations validate the tightness of the derived lower bounds and demonstrate substantial performance gains over RF MIMO, highlighting RAQ-MIMO satellites as a promising enabler for next-generation satellite communications.
\end{itemize}


The rest of the paper is organized as follows. The system model is presented in Section II. The channel estimation and performance analysis are provided in Section III. In Section IV, extensive numerical results are presented. Finally, the conclusions are drawn in Section V.

\emph{Notations}: \(\hbar\) is the reduced Planck constant and \(j^2 = -1\). \(\epsilon_0\) and \(c\) denote the vacuum permittivity and speed of light, respectively.  \(q\) and \(a_0\) respectively represent the elementary charge and Bohr radius. The superscripts \((\cdot)^T\), \((\cdot)^H\), and \((\cdot)^*\) denote transpose, Hermitian (conjugate transpose), and complex conjugation. $\mathrm{diag}(\cdot)$ denotes the diagonal matrix formed from a vector and \(\mathbf{I}_M\) means the \(M \times M\) identity. \([\mathbf{X}]_{m,n}\) denotes the \((m,n)\)-th element of \(\mathbf{X}\). \(\otimes\) is the Kronecker product.
 $\mathbb{C}$ represents a complex field and  $\mathcal{CN}(\boldsymbol{\mu},\boldsymbol{\Sigma})$  is a circularly symmetric complex Gaussian distribution with mean $\boldsymbol{\mu}$ and covariance $\boldsymbol{\Sigma}$. $\mathcal{CW}_m,(n,\boldsymbol{\Sigma},\boldsymbol{\Omega})$ is the complex non-central Wishart distribution with \(m\) dimension, \(n\) degrees of freedom, variance of \(\boldsymbol{\Sigma}\), and the non-centrality parameter \(\boldsymbol{\Omega}\).
$\mathbb{E}\{\cdot\}$ denotes expectation.

\section{System Model of RAQ-MIMO}
In this section, we briefly introduce the principles of RAQ-based receivers and then present the system model of the RAQ–MIMO.

\begin{figure*}
	\centering
	\includegraphics[width=7in]{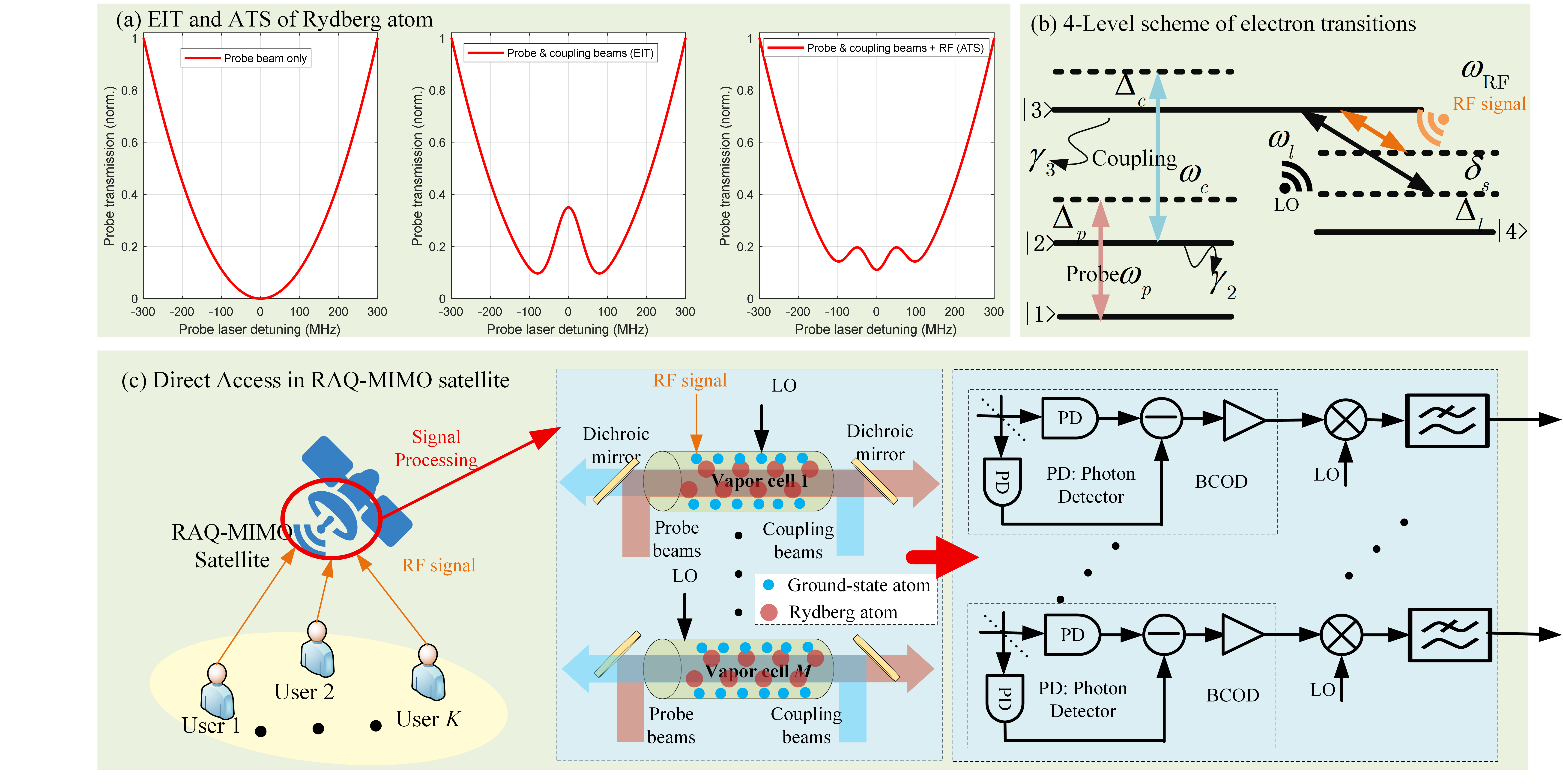}
	\caption{(a) Rydberg-EIT spectroscopy signal and ATS; (b) 4-Level scheme of electron transitions; (c) RAQ-MIMO satellite network, where the \(M\) atomic cells receive the signal from \(K\) users.}
	\label{system}
\end{figure*}

\subsection{Fundamental Principles}
Rydberg atoms provide a quantum-electrometric mechanism to detect RF signals by converting field-induced level dressing into an optical readout. As shown in Fig. \ref{system}(b), in a ladder configuration, a weak \emph{probe} laser resonantly drives the transition \(\lvert 1\rangle \!\to\! \lvert 2\rangle\), while a strong \emph{coupling} beam excites \(\lvert 2\rangle \!\to\! \lvert 3\rangle\), where \(\lvert 1\rangle \), \(\lvert 2\rangle \), and \(\lvert 3\rangle \) denote the ground, excited, and Rydberg state, respectively. Under two-photon resonance, quantum interference gives rise to EIT, as illustrated in Fig. \ref{system}(a), which suppresses absorption and creates a narrow transparency window in the probe transmission. 

When an external RF field couples a Rydberg state \(\lvert 4\rangle\) to a nearby Rydberg level \(\lvert 3\rangle\), it induces level dressing of the upper manifold and perturbs the EIT resonance. In the weak-to-intermediate RF regime, the applied RF field broadens and shifts the EIT resonance, whereas in the strong-field regime, the coupling gives rise to ATS of the EIT peak, as depicted in Fig. \ref{system}(a). The frequency separation \(f\) between the split peaks is directly proportional to the RF Rabi frequency, i.e.,
\begin{equation}\label{Rabi}
 f \;=\; \frac{\Omega_{\mathrm{RF}}}{2\pi}, \qquad \text{and}\quad
\Omega_{\mathrm{RF}} \;=\; \frac{\nu\, E_{\mathrm{RF}}}{\hbar},
\end{equation}
where \(\nu\) is the dipole matrix element and \(E_{\mathrm{RF}}\) is the RF electric-field amplitude. Consequently, the RF amplitude can be mapped onto an optical frequency scale with atomic-level intrinsic traceability. Frequency selectivity arises from the narrow \(\lvert 3\rangle\!\leftrightarrow\!\lvert 4\rangle\) transition, while vector information such as polarization can be inferred from atomic selection rules and angular dependence. By continuously monitoring the probe transmission (e.g., via photodetection), the RF amplitude, frequency, and phase can be recovered, thereby exploiting the large Rydberg dipole moments for ultra-high sensitivity.

\subsection{System model of RAQ-MIMO}
We begin with a description of RAQ-based signal processing. Assuming that the \(k\)-th user signal is a plane wave, the received signal of the \(m\)-th atomic sensor can be expressed as
\begin{equation}
	s_{m,k}(t) = \sqrt{2P_{m,k}}\cos(2\pi f_c t + \theta_{m,k}),
\end{equation}
where \(f_c\) denotes the carrier frequency, \(P_{m,k}\) represents the received power of user \(k\), and \(\theta_{m,k}\) is its corresponding phase. 

Similarly, we assume that the LO is also a plane wave, the received LO component at the \(m\)-th atomic sensor is given by
\begin{equation}
	s_{m,l}(t) = \sqrt{2P_{m,l}}\cos(2\pi f_l t + \theta_{m,l}),
\end{equation}
where \(f_l\), \(P_{m,l}\), and \(\theta_{m,l}\) denote the carrier frequency, received power, and phase of the LO signal, respectively.

The superimposed RF signal at the \(m\)-th sensor is then
\begin{equation}
	\begin{split}
		y_m(t) &= s_{m,l}(t) + \sum_{k = 1}^{K}s_{m,k}(t)\\
					&\overset{a}{\approx} \sqrt{2P_{y_m}}\cos(2\pi f_l t + \theta_{m,l}),
	\end{split}
\end{equation}
where the approximation \((a)\) follows from Appendix~A in \cite{gong2025rydberguplink}, and \(P_{y_m}\) denotes the amplitude of the superimposed signal, expressed as
\begin{equation}
	\begin{split}
			P_{y_m} =& P_{m,l} + \sum_{k_1=1}^{K}\sum_{k_2=1}^{K}\sqrt{P_{m,k_1}P_{m,k_2}}\\
			+&2\sum_{k=1}^K\sqrt{P_{m,l}P_{m,k}}\cos(2\pi \Delta f t + \Delta\theta_{m,k}),\\
	\end{split}
\end{equation} 
where \(\Delta f = f_c - f_l\) is the frequency offset and \(\Delta \theta_{m,k} = \theta_{m,k} - \theta_{m,l}\) represents the phase difference between the user and LO signals.

Based on the received signal at the \(m\)-th atomic sensor and (\ref{Rabi}), the Rabi frequency of the superimposed signal can be expressed as
\begin{equation}
	\Omega_{\mathrm{RF},m} \approx \Omega_{m,l} + \sum_{k=1}^{K}\Omega_{m,k}\cos(2\pi\Delta ft+\Delta\theta_{m,k}),
\end{equation}
where \(\Omega_{m,l}\) and \(\Omega_{m,k}\) denote the Rabi frequency corresponding to the LO and the \(k\)-th user's RF signal, respectively \footnote{Since the LO field is designed to be located near the atomic sensor, it satisfies \(\Omega_{m,l} \gg \Omega_{m,k}\)}. 

By employing the balanced coherent optical detection (BCOD) scheme and subsequent down-conversion, the time-varying component introduced by the \(K\) users can be detected. The sampled received signal is then expressed as \cite{gong2025rydberguplink,gong2024rydberg}
\begin{equation}
	 y_{m}(n) = \sqrt{\rho_m}\Phi_m \sum\limits_{k= 1}^K s_{m,k}(n) + w_m(n),
\end{equation}
where \(\rho_m\) and \(\Phi_m\) denote the effective gain and phase shift of the \(m\)-th vapor cell, respectively \footnote{The detailed derivation can be found in \cite{gong2025rydberguplink,gong2024rydberg}.}. The noise term \(w_m\) follows a complex Gaussian distribution \(\mathcal{CN}(0,\sigma^2)\), where \(\sigma^2 = \frac{N_{\text{QPN}}+N_{\text{PSN}}+N_{\text{ITN}}}{2}\) is the summed power of the quantum projection noise, PD shot noise, and intrinsic thermal noise \cite{gong2024rydberg}.

By appropriately configuring the system parameters, the vapor cells can be calibrated to yield identical gain. The resulting RAQ–MIMO system model is expressed as \cite{gong2024rydberg,gong2025rydberguplink}
\begin{equation}
    \label{systemmodel}
    \small
    \mathbf{y} = \sqrt{\rho}\Phi\mathbf{D}\mathbf{H}\mathbf{s} + \mathbf{w},
\end{equation}
where \(\mathbf{H} = [\mathbf{h}_1,\cdots,\mathbf{h}_K] \in \mathbb{C}^{M \times K}\) denotes the channel matrix, \( \mathbf{w} \sim \mathcal{CN}(\mathbf{0},\sigma^2\mathbf{I}_M)\) denotes additive white Gaussian noise (AWGN), \(\Phi \triangleq \frac{e^{-j(\theta_l - \varphi(\Omega_l))}}{2} +\frac{e^{-j(\theta_l + \varphi(\Omega_l))}}{2} \) captures the phase shift determined by the first vapor cell's phase \(\theta_{l} \triangleq \theta_{1,l} \) and the superposition-induced phase term \(\varphi(\Omega_l)\) (see eq. (12) in \cite{gong2025rydberguplink}), and \(\rho\triangleq \rho_m, \forall m\) is the effective gain. The matrix \(\mathbf{D} = \text{diag}\{1,e^{-j\frac{2\pi d\sin\vartheta}{\lambda}},\cdots, e^{-j\frac{2\pi (M-1)d\sin\vartheta}{\lambda}}\}\) represents the phase shift across the array, where \(\vartheta\) is the angle of arrival of the LO signal, \(d\) is the inter-element distance between vapor cells, and \(\lambda\) is the wavelength. 

Assuming a Rician fading channel between the users and satellite, the channel vector of the \(k\)-th user is modeled as
\begin{equation}
    \mathbf{h}_k = \sqrt{\frac{\beta_k\delta_k}{\delta_k+1}} \mathbf{\bar h}_k + \sqrt{\frac{\beta_k}{\delta_k+1}} \mathbf{{\tilde h}}_k,
\end{equation}
where \( \delta_k\) denotes the Rician factor and \(\beta_k\) the large-scale fading factor. The term \(\mathbf{{\tilde h}}_k\) represents the non-line-of-sight (NLoS) component following the complex Gaussian distribution with zero mean and unit variance, whereas \(\mathbf{\bar h}_k\) denotes the LoS component with respect to elevation angle \(\theta^{e}_{k}\) and azimuth angle \(\phi_{k}\) from the satellite to user \(k\), expressed as
\begin{equation}
	\mathbf{\bar h}_k = [1,e^{-j2\pi\frac{d}{\lambda}\sin\theta^{e}_{k} \cos\phi_{k}},\cdots,e^{-j2\pi(M-1)\frac{d}{\lambda}\sin\theta^{e}_{k} \cos\phi_{k}}]^T,
\end{equation}
where \(\lambda\) is the wavelength of the carrier frequency.

\section{Performance Analysis under Imperfect CSI}
This section presents a comprehensive performance analysis of the RAQ–MIMO system under imperfect CSI, focusing on the closed-form derivation of the MSE and the achievable rate lower bound.

\subsection{Channel Estimation}
Based on the system model in (\ref{systemmodel}), the received signal at the atomic receiver is given by
\begin{equation}
    \label{pilot}
    \small
    \mathbf{Y} = \sqrt{\rho}\Phi\mathbf{D}\mathbf{H}\mathbf{P}\mathbf{Q}^H + \mathbf{W},
\end{equation}
where \(\mathbf{W}\) is the background noise matrix, \(\mathbf{P} = \text{diag}\{\sqrt{\tau p^p_1},\cdots,\sqrt{\tau p^p_K}\}\) represents the pilot energy matrix determined by the pilot power \(p^p_k\) and pilot length \(\tau\), and \(\mathbf{Q} = [\mathbf{q}_1,\cdots,\mathbf{q}_K]  \in \mathbb{C}^{\tau \times K}\) denotes the pilot sequence with \(\mathbf{q}^H_k\mathbf{q}_{k'} = 0\) for \(k \neq k'\) or \(\mathbf{q}^H_k\mathbf{q}_{k'} = 1\) for \(k = k'\). By correlating (\ref{pilot}) with \(\mathbf{q}_k\), we obtain
\begin{equation}
    \label{rxpilot}
    \small
    \mathbf{y}_k =  \sqrt{\rho \tau p^p_k}\Phi\mathbf{D}\mathbf{h}_k + \mathbf{W}\mathbf{q}_k.
\end{equation}

The minimum mean-square-error (MMSE) estimate of \(\mathbf{h}_k\) is expressed as
\begin{equation}
    \label{estimatedchannel}
    \small
    \begin{split}
          \mathbf{\hat h}_k &= \sqrt{\frac{\beta_k \delta_k}{\delta_k + 1}}\mathbf{\bar h}_k +\text{Cov}\Big\{\sqrt{\frac{\beta_k}{\delta_k + 1}}\mathbf{\tilde h}_k,\mathbf{y}_k\Big\}\text{Cov}^{-1}\{\mathbf{y}_k,\mathbf{y}_k\}\\
          &\times\Big(\mathbf{y}_k - \mathbb{E}\{\mathbf{y}_k\}\Big) \\
          & = \sqrt{\frac{\beta_k \delta_k}{\delta_k + 1}}\mathbf{\bar h}_k  + \sqrt{\rho \tau p^p_k}{\frac{\beta_k}{\delta_k + 1}}\Phi^H \mathbf{D}^H\\
           &\times\Big(\frac{\rho \tau p^p_k \beta_k |\Phi|^2}{\delta_k + 1 }+ \sigma^2\Big)^{-1}
          \Big(\sqrt{\frac{\rho \tau p^p_k\beta_k}{\delta_k +1}}\Phi\mathbf{D}\mathbf{\tilde h}_k+\mathbf{Wq}_k\Big).
    \end{split}
\end{equation}
By defining \(\alpha_k\triangleq\frac{\beta_k}{\delta_k +1}\), the estimated channel can be rewritten as
\begin{equation}
\begin{split}
        \mathbf{\hat h}_k &= \sqrt{\delta_k \alpha_k} \mathbf{\bar h}_k + \frac{\rho \tau p^p_k \alpha^{3/2}_k |\Phi|^2}{\rho \tau p^p_k \alpha_k |\Phi|^2 + \sigma^2}\mathbf{\tilde h}_k \\
        &+ \frac{\sqrt{\rho \tau p^p_k }\alpha_k \Phi^H}{\rho \tau p^p_k \alpha_k |\Phi|^2 + \sigma^2}\mathbf{D}^H\mathbf{Wq}_k.
\end{split}
\end{equation}
Accordingly, the estimated channel follows the complex distribution \( \mathbf{\hat h}_k \sim \mathcal{CN}(\sqrt{\delta_k \alpha_k} \mathbf{\bar h}_k, \frac{\rho \tau p^p_k \alpha^2_k |\Phi|^2}{\rho \tau p^p_k c_k |\Phi|^2 + \sigma^2}\mathbf{I}_{M})\).  By denoting the estimation error as \(\mathbf{e}_k = \mathbf{h}_k-\mathbf{\hat h}_k\), the MSE matrix is given by
\begin{equation}
    \label{MSE}
    \small
    \mathbf{MSE}_k = \frac{\alpha_k\sigma^2}{\rho \tau p^p_k\alpha_k|\Phi|^2 + \sigma^2} \mathbf{I}_M \triangleq( \alpha_k -\mu_k\alpha_k )\mathbf{I}_M,
\end{equation}
where \(\mu_k\) is \(\frac{\rho\tau p^p_k\alpha_k |\Phi|^2}{\rho\tau p^p_k\alpha_k |\Phi|^2+\sigma^2}\). The NMSE of user \(k\) is therefore
\begin{equation}
    \text{NMSE}_k = \frac{\sigma^2}{\rho\tau p^p_k\alpha_k|\Phi|^2 + \sigma^2} = 1-\mu_k.
\end{equation}

By denoting equivalent signal-to-noise ratio (SNR) as \(\gamma_k \triangleq \frac{\rho p^p_k \beta_k |\Phi|^2}{\sigma^2} \), we have \(\text{MSE}_k = \frac{M \beta_k}{1+\tau \gamma_k}\) when \(\delta_k \rightarrow 0\). This result confirms that RAQ–MIMO and conventional MIMO systems exhibit similar channel estimation behavior and that the accuracy of channel estimation depends on both channel quality and pilot length. However, owing to the quantum-enhanced characteristics of RAQRs, additional estimation gains can be achieved, as summarized below.

\begin{corollary}
    \emph{Compared with conventional RF MIMO, the channel estimation in RAQ–MIMO can be enhanced by a factor of \(10\lg\Big(\frac{\rho|\Phi|^2/\sigma^2}{\rho_{\text{RF}} /\sigma^2_{\text{RF}}}\Big)\) in the high SNR regime, where \(\rho_{\text{RF}}\) and \(\sigma^2_{\text{RF}}\) denote the effective gain and noise background of RF MIMO, respectively}.
    
   \emph{Proof}: The proof is omitted for brevity. $\hfill\blacksquare$
\end{corollary}

\begin{remark}[Breaking the 3-dB Barrier via a Power-Free Mechanism]
In a conventional MIMO system, doubling the estimation accuracy typically requires doubling the pilot power or the pilot length~\(\tau\). In contrast, because \(\gamma_k \propto |\Phi|^2 \) or \(\gamma_k \propto \rho \) in RAQRs, the same improvement can be achieved by increasing \(\Phi\) by a factor of \(\sqrt{2}\) or by scaling \(\rho\) accordingly, without extra power consumption (assuming \(\sigma^2\) remains constant) \footnote{Recent studies on Rydberg receivers demonstrate practical approaches to increasing \(\Phi\) without proportionally raising the noise back ground \(\sigma^2\), for example by enhancing the EIT signal slope \cite{gong2025rydberguplink}.}.
\end{remark}

\begin{remark}[Existence of an Optimal \(\Phi^{\star}\)] From (\ref{MSE}), the channel estimation error is minimized when \(\Phi\) is maximized. Since \(\Phi\) is a function of the LO phase at the first atomic sensor~\cite{gong2025rydberg}, the optimal \(\Phi^{\star} \) can be obtained through proper design of the LO phase.
\end{remark}

     

\subsection{Performance Analysis}
Based on the estimated channel \(\mathbf{\hat h}_k\), the receiver performs linear detection, such as MRC and ZF. Let \(\mathbf{c}_k\) the linear detection vector, and assume that the receiver has knowledge of the statistical CSI. The detected signal for user $k$ can be expressed as \cite{peng2022resource}
\begin{equation}
    \label{rxsignal}
    \small
    \begin{split}
         r_k& = \underbrace{ \mathbb{E}\{\sqrt{\rho p^d_k}\Phi\mathbf{c}^{H}_k\mathbf{D}\mathbf{h}_k\}s_k}_{\text{Ds}_k} \\
        &+ \underbrace{ \sqrt{\rho p^d_k}\Phi\mathbf{c}^{H}_k\mathbf{D}\mathbf{h}_ks_k -\mathbb{E}\{\sqrt{\rho p^d_k}\Phi\mathbf{c}^{H}_k\mathbf{D}\mathbf{h}_k\}s_k}_{\text{Ls}_k}  \\
        & + \underbrace{\sum\limits_{k' \neq k}^K\sqrt{\rho p^d_{k'}}\Phi\mathbf{c}^{H}_k\mathbf{D}\mathbf{h}_{k'}s_{k'}}_{\text{UI}_{k,k'}}+\underbrace{\mathbf{c}^{H}_k\mathbf{w} }_{\text{N}_k},
    \end{split}
\end{equation}
where \(s_k \sim \mathcal{CN}(0,1)\) is the mutually independent data symbol, \(p^d_k\), \(\text{Ds}_k\), \(\text{Ls}_k\), \(\text{UI}_{k,k'}\), and \(\text{N}_k\) represent the \(k\)-th user's transmit power, effective signal, leaked signal, inter-user interference, and noise, respectively. Accodingly, the achievable data rate \(R_k\) is given by
\begin{equation}
    \label{achRate}
    R_k = \frac{T-\tau}{T}\log_2\Big(1+\frac{|\text{Ds}_k|^2}{|\text{Ls}_k|^2 + \sum\limits_{k' \neq k}^K|\text{UI}_{k,k'}|^2 + |\text{N}_k|^2}\Big),
\end{equation}
where \(T\) is the coherence interval (in symbols).

Since deriving the closed-form expression of (\ref{achRate}) is intractable, a tight analytical lower bound for the achievable rate of user \(k\) is derived in the following theorems.

\begin{theorem}
\label{MRC_SINR_T}
The ergodic achievable rate for the $k$-th user employing the MRC decoder is lower bounded as
\begin{equation}
\setlength\abovedisplayskip{5pt}
\setlength\belowdisplayskip{5pt}
\small
\label{MRC_LB_rate}
R_k \ge {\underline R}_k^{\rm MRC} = \frac{T-\tau}{T}\log_2(1+\text{SINR}^{\text{MRC}}_k),
\end{equation}
where \(\text{SINR}^{\text{MRC}}_k\) is defined in (\ref{SINR_MRC}), given at the bottom of this page.

\begin{figure*}[b]
      \hrule
    \centering
   \begin{equation}\label{SINR_MRC}
        \text{SINR}^{\text{MRC}}_k = \frac{Mp^d_k\alpha_k(\mu_k+\delta_k)^2}{\sum\limits_{k' = 1}^K p^d_{k'}(\delta_k\alpha_{k'}+\mu_k\beta_{k'})+\sum\limits_{k' \neq k}^K p^d_{k'} \frac{|\mathbf{\bar h}^H_k\mathbf{\bar h}_{k'}|^2}{M}\delta_k\delta_{k'}\alpha_{k'} + \frac{\sigma^2}{\rho |\Phi|^2}(\mu_k + \delta_k)}.
   \end{equation}
\end{figure*}


\emph{Proof}: Please refer to Appendix \ref{Prooftheorem1}. $\hfill\blacksquare$
\end{theorem}

\begin{theorem}
\label{ZF_SINR_T}
The ergodic achievable rate for the $k$-th user employing the ZF decoder is lower bounded as
\begin{equation}
\setlength\abovedisplayskip{5pt}
\setlength\belowdisplayskip{5pt}
\small
\label{ZF_LB_rate}
R_k \ge {\underline R}_k^{\rm ZF} = \frac{T-\tau}{T}\log_2(1+\text{SINR}^{\text{ZF}}_k),
\end{equation}
where \(\text{SINR}^{\text{ZF}}_k\) is given by
\begin{equation}
   \label{ZF_up_SINR}
   \begin{split}
          &\text{SINR}^{\text{ZF}}_k \\
          =& \frac{(M-K) p^d_k }{( \sum\limits_{k' = 1}^K p^d_{k'}(\alpha_{k'} - \mu_{k'}\alpha_{k'}) + \frac{\sigma^2}{|\Phi|^2 \rho})\big[(\boldsymbol{\Psi}+\frac{\mathbf{\bar H}^H\mathbf{\bar H}}{M})^{-1}\big]_{k,k}},
   \end{split}
\end{equation}
where \([\mathbf{X}]_{k,k}\) denotes the \((k,k)\)-th element of matrix \(\mathbf{X}\), \(\boldsymbol{\Psi} = \text{Diag}\{\mu_1\alpha_1,\cdots,\mu_K\alpha_K\}\) is the diagonal matrix of channel estimation gains, and \(\mathbf{\bar H} = [\sqrt{\delta_1\alpha_1}\mathbf{\bar h}_1,\cdots,\sqrt{\delta_K\alpha_K}\mathbf{\bar h}_K]\) collects the LoS channel components.

\emph{Proof}: Please refer to Appendix \ref{Prooftheorem2}. $\hfill\blacksquare$
\end{theorem}

The closed-form expressions derived in Theorems~\ref{MRC_SINR_T} and~\ref{ZF_SINR_T} depend only on the large-scale fading factors and the key parameters RAQRs. Compared with time-consuming simulation-based evaluations, these analytical expressions offer valuable insights for the design and resource allocation of RAQ–MIMO systems.

In the following, we examine the impact of imperfect CSI on the derived lower bounds under three representative channel conditions: 1) Rayleigh fading, 2) pure LoS, and 3) typical satellite channels.

\subsubsection{Rayleigh Channel}

From Theorem~\ref{MRC_SINR_T}, the MRC bound under Rayleigh fading can be expressed as
\begin{equation}
\mathrm{SINR}_{k,\text{Ray}}^{\rm MRC}
=\underbrace{\frac{M\,p_k^d\,\beta_k}{\sum_{k'} p_{k'}^d\beta_{k'} + \sigma^2/(|\Phi|^2\rho)}}_{\text{perfect-CSI baseline (RAQR-MRC)}}
\times \underbrace{\frac{\rho\,\tau\,p_k^p\,\beta_k\,|\Phi|^2}{\rho\,\tau\,p_k^p\,\beta_k\,|\Phi|^2+\sigma^2}}_{\textstyle 1-\mu_k}.
\label{eq:MRC-factor}
\end{equation}
Hence, the overall performance degradation due to imperfect CSI in MRC is captured by the multiplicative factor \((1 - \mu_k)\).

For ZF detection, (\ref{ZF_up_SINR}) can be rewritten as
\begin{equation}
\begin{split}
\mathrm{SINR}_{k,\text{Ray}}^{\rm ZF}
&=\underbrace{\frac{(M-K)\,p_k^d\,\beta_k}{\sigma^2/(|\Phi|^2\rho)}}_{\text{perfect-CSI baseline (RAQR-ZF)}} \times \underbrace{\frac{\rho\,\tau\,p_k^p\,\beta_k\,|\Phi|^2}{\rho\,\tau\,p_k^p\,\beta_k\,|\Phi|^2+\sigma^2}}_{\textstyle 1-\mathrm{NMSE}_k} \\
&\times \frac{1}{\,1+\Delta_{\rm RI}\,}.
\end{split}
\label{eq:ZF-factor}
\end{equation}
where \(\Delta_{\mathrm{RI}}\) represents the residual interference resulting from imperfect CSI, defined as
\begin{equation}
\label{delta}
\Delta_{\rm RI} \triangleq \sum\limits_{k'}  \frac{p^d_{k'}\beta_{k'}}{\tau p^{p}_{k'}\beta_{k'}+\sigma^2/(|\Phi|^2\rho)}.
\end{equation}

\begin{remark}\label{rem:imperfect-vs-perfect}
\emph{For the Rayleigh fading channel, imperfect CSI introduces a multiplicative loss of $(1-\mathrm{NMSE}_k)$ for MRC and  $(1-\mathrm{NMSE}_k) \times \frac{1}{\,1+\Delta_{\rm RI}}$ for ZF. Therefore, ZF detection benefits more from the improved channel estimation than MRC.}
\end{remark}

\subsubsection{LoS Channel}

When the Rician factor \(\delta_k \rightarrow \infty, \forall k\), the lower bounds for MRC and ZF can be respectively written as
\begin{equation}
    \mathrm{SINR}_{k,\text{LoS}}^{\rm MRC}
= \frac{Mp^d_k \beta_k}{\sum\limits_{k' \neq k}^K p^d_{k'}\frac{|\mathbf{\bar h}^H_k\mathbf{\bar h}_{k'}|^2}{M}\beta_{k'} + \frac{\sigma^2}{\rho |\Phi|^2} },
\end{equation}
and 
\begin{equation}
    \mathrm{SINR}_{k,\text{LoS}}^{\rm ZF}
= \frac{(M-K)p^d_k }{\frac{\sigma^2}{\rho |\Phi|^2} \Big[(\frac{\mathbf{\bar H}^H\mathbf{\bar H}}{M})^{-1}\Big]_{k,k} }.
\end{equation}

Since the channels between users and RAQ-MIMO are deterministic under pure LoS conditions, no performance loss arises from channel estimation errors.

\subsubsection{Typical Satellite Channel}
Owing to dominant LoS propagation, the Rician factor, \(\delta_k\) in satellite links typically lies within \([10,100]\). Thus, the user–satellite channels are nearly deterministic, and channel estimation errors become negligible. Consequently, based on the derivation of \eqref{eq:MRC-factor} and \eqref{eq:ZF-factor}, the performance degradation due to imperfect CSI is marginal.

\begin{remark}[Power Scaling Law]\label{rem:scaling}
\emph{For both Rayleigh and Rician fading channels, the power scaling law continues to hold for multi-user RAQRs.}

Assume that each user's transmit power satisfies
\begin{equation}
p_k^d=\frac{E}{M^{\epsilon_d}},\qquad
\tau p_k^p=\frac{E}{M^{\epsilon_p}},\qquad E>0.
\label{eq:scaling-RAQR}
\end{equation}
As \(M \to \infty\), we have
\begin{equation}
    \begin{split}
        &\frac{|\mathbf{\bar h}^H_k\mathbf{\bar h}_{k'}|^2}{M} \rightarrow \mathcal{O}(1), \mu_k \rightarrow 0.
    \end{split}
\end{equation}
Hence, for fixed $K$, $\beta_k$, and $(\rho,\Phi,\sigma^2)$, the asymptotic power-scaling law for the RAQ–MIMO system can be expressed as
\begin{equation}
    \mathrm{SINR}_k^{\rm MRC/ZF} \xrightarrow{M\to\infty} \begin{cases}
\log_2\Big(1+\frac{\rho^2|\Phi|^4\beta^2_kE^2}{\sigma^4}\Big), & \delta_k = 0\\
\log_2\Big(1+\frac{\rho|\Phi|^2\delta_k\beta_kE}{(\delta_k + 1)\sigma^2}\Big), & \delta_k \neq 0.
\end{cases}
\end{equation}

Note that the non-vanishing rate is achieved only under the following two cases:
\begin{enumerate}
    \item \textbf{Case I}: For \(\delta_k = 0\), the non-zero rate occurs only when \(\epsilon_d + \epsilon_p = 1\);
    \item \textbf{Case II}: For \(\delta_k \neq 0\), the non-vanishing rate exists only when \(\epsilon_d = 1\);
\end{enumerate}

These results reproduce the classical power-scaling behaviors reported in~\cite{ngo2013energy}: perfect-CSI laws ($p^d_k\!\propto\!1/M$) and imperfect CSI laws ($p^d_k,p^p_k\!\propto\!1/\sqrt{M}$) for both MRC and ZF. Furthermore, for Rayleigh fading channels, the transmit power can be reduced by a factor of \(\frac{1}{\sqrt{M}}\), whereas in pure LoS channels it can be reduced by \(\frac{1}{M}\), owing to reduced fading fluctuations and a higher received SINR. More importantly, the RAQR does not alter the power-scaling exponent but enhances the non-vanishing rate through the multiplicative gain $\rho|\Phi|^2$.
\end{remark}

\subsection{Benefits of RAQ-MIMO over Conventional MIMO}

For conventional MIMO, the parameters are given by \(\Phi =\Phi_{\text{RF}}= 1\), \(\rho = \rho_{\text{RF}}= A_{\text{iso}}\eta_0 G_{\text{Ant}}G_{\text{LNA}}\), and \(\sigma^2 =\sigma^2_{\text{RF}} = 10\lg(k_{B}T_0) +10\lg(B)+10\lg(F)+G_{\text{LNA}} \), where \(A_{\text{iso}} = \lambda^2/(4\pi)\) is the effective aperture of an isotropic antenna with respect to wavelength \(\lambda\), \(\eta_0\) denotes the antenna efficiency, \(G_{\text{Ant}}\) represents the antenna gain, and \(G_{\text{LNA}}\) is the low-noise amplifier (LNA) gain, \(k_{B}\) is the Boltzmann constant, \(T_0\) denotes the temperature, \(B\) represents the bandwidth, and \(F\) is the noise factor. Accordingly, the lower bounds for the achievable rates of RF MIMO systems are expressed as (\ref{MRCRF}) and (\ref{ZFRF}), given at the bottom of the next page. The channel-estimation gain coefficient in~(\ref{MRCRF}) is expressed as \(\mu_{k,\text{RF}} =\frac{\rho_{\text{RF} }\tau p^p_k \alpha_k}{\rho_{\text{RF} }\tau p^p_k \alpha_k + \sigma^2_{\text{RF}}}\), and the diagonal matrix  \(\boldsymbol{\Psi}_{\text{RF}} = \text{Diag}\{\mu_{1,\text{RF}},\cdots,\mu_{K,\text{RF}}\}\) in~(\ref{ZFRF}) collects the channel-estimation gains for all users.
\begin{figure*}[b]
\hrule
\begin{equation}\label{MRCRF}
    \mathrm{SINR}_{k,\text{RF}}^{\rm MRC} = \frac{Mp^d_k\alpha_{k}(\mu_{k,\text{RF}}+\delta_k)^2}{\sum\limits_{k' = 1}^K p^d_{k'}(\delta_k\alpha_{k'}+\mu_{k,\text{RF}}\beta_{k'})+\sum\limits_{k' \neq k}^K p^d_{k'} \frac{|\mathbf{\bar h}^H_k\mathbf{\bar h}_{k'}|^2}{M}\delta_k\delta_{k'}\alpha_{k'} + \frac{\sigma^2_{k,\text{RF}}}{\rho_{\text{RF}}}(\mu_{k,\text{RF}} + \delta_k)}.
\end{equation}    
\hrule
\begin{equation} \label{ZFRF}
     \text{SINR}^{\text{ZF}}_{k,\text{RF}} 
          = \frac{(M-K) p^d_k }{( \sum\limits_{k' = 1}^K p^d_{k'}(\alpha_{k'} - \mu_{k',\text{RF}}\alpha_{k'}) + \frac{\sigma^2_{\text{RF}}}{ \rho_{\text{RF}}})\big[(\boldsymbol{\Psi}_{\text{RF}}+\frac{\mathbf{\bar H}^H\mathbf{\bar H}}{M})^{-1}\big]_{k,k}}.
\end{equation}
\end{figure*}

Next, we rewrite the lower bounds of RAQ-MIMO as (\ref{MRCRAQ}) and (\ref{ZFRAQ}), given at the bottom of the next page. We then evaluate the lower bounds to quantify the performance gains of RAQ-MIMO relative to RF MIMO.

\begin{remark}[Advantages of RAQR over Conventional MIMO]\label{rem:RAQR-vs-trad}
\emph{The advantages of RAQR originate from two \textbf{power-free} tuning parameters} ($\Phi,\rho$) \emph{as well as from a reduced noise background. }

As revealed in (\ref{MRCRAQ}) and (\ref{ZFRAQ}), the performance gain of RAQ-MIMO stems from the improved channel estimation accuracy and normalized noise. Compared with conventional receivers, RAQR introduces two controllable degrees of freedom, including
$\Phi$ and $\rho$.
 Both MRC and ZF benefit from the multiplicative gain factor \(\rho |\Phi|^2\), which enhances the signal-to-noise ratio without increasing transmit power; (ii) The effective noise term \(\sigma^{2}/(|\Phi|^{2}\rho)\) is reduced, thereby improving the perfect-CSI baseline performance for both MRC and ZF receivers. Furthermore, the gain of RAQR can be maximized by configuring \(\Phi\), consistent with the findings in Remark~3 of~\cite{gong2025rydberguplink}.
\end{remark}

\begin{figure*}[b]
\hrule
\begin{equation}\label{MRCRAQ}
    \mathrm{SINR}_{k}^{\rm MRC} = \mathrm{SINR}_{k,\text{RF}}^{\rm MRC} \times (\frac{\mu_k+\delta_k}{\mu_{k,\text{RF}}+\delta_k})^2 \times \frac{\sum\limits_{k' = 1}^K p^d_{k'}(\delta_k\alpha_{k'}+\mu_{k,\text{RF}}\beta_{k'})+\sum\limits_{k' \neq k}^K p^d_{k'} \frac{|\mathbf{\bar h}^H_k\mathbf{\bar h}_{k'}|^2}{M}\delta_k\delta_{k'}\alpha_{k'} + \frac{\sigma^2_{\text{RF}}}{\rho_{\text{RF}}}(\mu_{k,\text{RF}} + \delta_k)}{\sum\limits_{k' = 1}^K p^d_{k'}(\delta_k\alpha_{k'}+\mu_{k}\beta_{k'})+\sum\limits_{k' \neq k}^K p^d_{k'} \frac{|\mathbf{\bar h}^H_k\mathbf{\bar h}_{k'}|^2}{M}\delta_k\delta_{k'}\alpha_{k'} + \frac{\sigma^2}{\rho |\Phi|^2}(\mu_{k} + \delta_k)} .
\end{equation}    
\hrule
\begin{equation} \label{ZFRAQ}
     \text{SINR}^{\text{ZF}}_{k} 
          = \text{SINR}^{\text{ZF}}_{k,\text{RF}} \times \frac{ \sum\limits_{k' = 1}^K p^d_{k'}(\alpha_{k'} - \mu_{k',\text{RF}}\alpha_{k'}) + \frac{\sigma^2_{\text{RF}}}{ \rho_{\text{RF}}}}{ \sum\limits_{k' = 1}^K p^d_{k'}(\alpha_{k'} - \mu_{k'}\alpha_{k'}) + \frac{\sigma^2}{ \rho |\Phi|^2}} \times \frac{\big[(\boldsymbol{\Psi}_{\text{RF}}+\frac{\mathbf{\bar H}^H\mathbf{\bar H}}{M})^{-1}\big]_{k,k}}{\big[(\boldsymbol{\Psi}+\frac{\mathbf{\bar H}^H\mathbf{\bar H}}{M})^{-1}\big]_{k,k}}.
\end{equation}
\end{figure*}

In the following, we compare the performance of RAQ–MIMO and RF MIMO under Rayleigh and Rician fading conditions.

\subsubsection{Rayleigh Channel}
By substituting \(\delta_k = 0\) into (\ref{MRCRAQ}) and (\ref{ZFRAQ}), the SINR  for MRC and ZF detection can be expressed as
\begin{equation}
\begin{split}
        \mathrm{SINR}_{k,\text{Ray}}^{\rm MRC} &= \mathrm{SINR}_{k,\text{RF},\text{Ray}}^{\rm MRC}\times \frac{\sum_{k'}p^d_{k'}\beta_{k'} + \sigma^{2}_{\text{RF}}/\rho_{\text{RF}}}{\sum_{k'}p^d_{k'}\beta_{k'} + \sigma^{2}/(\rho|\Phi|^2)}\\
        &\times\frac{\tau p^p_k \beta_k + \sigma^2_{\text{RF}}/\rho_{\text{RF}}}{\tau p^p_k \beta_k + \sigma^2/(\rho|\Phi|^2)}, \\
\end{split}
\end{equation}
and 
\begin{equation}
\begin{split}
        \mathrm{SINR}_{k,\text{Ray}}^{\rm ZF} &= \mathrm{SINR}_{k,\text{RF},\text{Ray}}^{\rm ZF}\times \frac{\Delta_{\rm RI,RF}+1}{\Delta_{\rm RI}+1}\times\frac{\sigma^2_{\text{RF}}/\rho_{\text{RF}}}{\sigma^2/(\rho|\Phi|^2)}\\
        &\times \frac{\tau p^p_k \beta_k + \sigma^2_{\text{RF}}/\rho_{\text{RF}}}{\tau p^p_k \beta_k + \sigma^2/(\rho|\Phi|^2)}, \\
\end{split}
\end{equation}
where \(\Delta_{\rm RI,RF} = \sum\limits_{k'}  \frac{p^d_{k'}\beta_{k'}}{\tau p^{p}_{k'}\beta_{k'}+\sigma^2_{\text{RF}}/(\rho_{\text{RF}})}\) is the residual interference factor.

\begin{corollary}\label{gain}
\emph{In the high-SINR regime (\(p^d_k, p^p_k\rightarrow\infty\)), the achievable rate difference between RAQR and conventional RF MIMO systems is approximated as}
\begin{equation}
\begin{split}
    \Delta R^{\mathrm{MRC}}_{k,\text{Ray}}&\approx 0,   \\
    \Delta R^{\mathrm{ZF}}_{k,\text{Ray}}&\approx\frac{T-\tau}{T}\log_2\Big(\frac{\rho|\Phi|^2/\sigma^2}{\rho_{\text{RF}} /\sigma^2_{\text{RF}}}\Big).
\end{split}
\end{equation}

\emph{Proof}: The proof is omitted for brevity. $\hfill\blacksquare$
\end{corollary}



\begin{corollary}\label{gain3}
\emph{In the low-SINR regime}
\begin{equation}
\begin{split}
    \Delta R^{\mathrm{MRC}}_{k,\text{Ray}}&\approx \frac{T-\tau}{T}\log_2\Big(1+\mathrm{SINR}_{k,\text{RF}}^{\rm MRC}\frac{\rho^2|\Phi|^4}{\sigma^4}\Big),   \\
    \Delta R^{\mathrm{ZF}}_{k,\text{Ray}}&\approx\frac{T-\tau}{T}\log_2\Big(1+\mathrm{SINR}_{k,\text{RF}}^{\rm ZF}\frac{\rho^2|\Phi|^4}{\sigma^4}\Big).
\end{split}
\end{equation}

\emph{Proof}: The result follows from the first-order approximation \(\log_2(1+x) \!\approx\! x / \ln(2)\) for \(x \!\to\! 0\). $\hfill\blacksquare$
\end{corollary}

\begin{remark}[Squaring Effect]
    \emph{In contrast to the constant gain observed in the high-SINR regime, RAQRs exhibit a noticeable performance improvement in the low-power region owing to the \emph{``squaring effect''}, where the rate gain scales proportionally with the SINR.}
\end{remark}

\begin{corollary}\label{Powerray}
     \emph{For Rayleigh fading channels, the transmit power of users in RAQ-MIMO systems can be reduced by a factor of \(\frac{\rho|\Phi|^2/\sigma^2}{\rho_{\text{RF}}/\sigma^2_{\text{RF}}}\). Equivalently, for the same transmit power, RAQ-MIMO can receive radio frequency signals from a longer distance by a factor of $\sqrt{\frac{\rho|\Phi|^2/\sigma^2}{\rho_{\text{RF}}/\sigma^2_{\text{RF}}}}$, which is consistent with Corollaries~4 and~5 in~\cite{gong2025rydberguplink}. }
     
     \emph{Proof}: The result follows by equating the achievable SINRs, i.e., \(\mathrm{SINR}_{k,\text{RF},\text{Ray}}^{\rm MRC} = \mathrm{SINR}_{k,\text{Ray}}^{\rm MRC}\) or \(\mathrm{SINR}_{k,\text{RF},\text{Ray}}^{\rm ZF} = \mathrm{SINR}_{k,\text{Ray}}^{\rm ZF}\), which implies that the same data rate can be achieved with transmit power reduced by the factor \(\frac{\rho|\Phi|^2/\sigma^2}{\rho_{\text{RF}}/\sigma^2_{\text{RF}}}\). Since the free-space path loss scales quadratically with the propagation distance, this corresponds to a range-extension factor of \(\sqrt{\frac{\rho|\Phi|^2/\sigma^2}{\rho_{\text{RF}}/\sigma^2_{\text{RF}}}}\). $\hfill\blacksquare$
\end{corollary}
   
It is worth noting that this capability to substantially reduce the required transmit power can effectively mitigate link budget challenges associated with long-distance propagation, thereby offering a promising solution for ground–space direct access.

\begin{corollary}\label{antennasize}
\emph{For Rayleigh fading channels, the number of antennas in RAQ–MIMO systems can be reduced by a factor of \(\frac{\rho|\Phi|^2/\sigma^2}{\rho_{\text{RF}}/\sigma^2_{\text{RF}}}\) and even by \((\frac{\rho|\Phi|^2/\sigma^2}{\rho_{\text{RF}}/\sigma^2_{\text{RF}}})^2\) in low-power regime. }

\emph{Proof}: The result follows by equating the achievable SINRs, i.e., \(\mathrm{SINR}_{k,\text{RF},\text{Ray}}^{\rm MRC} = \mathrm{SINR}_{k,\text{Ray}}^{\rm MRC}\) or \(\mathrm{SINR}_{k,\text{RF},\text{Ray}}^{\rm ZF} = \mathrm{SINR}_{k,\text{Ray}}^{\rm ZF}\).  $\hfill\blacksquare$
\end{corollary}

Corollary \ref{antennasize} indicates that integrating RAQ-MIMO on satellites can substantially reduce both the number and the required aperture of antennas, thereby alleviating payload and launch constraints of next-generation small satellites.

\subsubsection{LoS Channel}
For the pure LoS case, i.e., when \(\delta_k \rightarrow \infty\), the lower bounds of RAQ-MIMO can be expressed as
\begin{equation}
     \mathrm{SINR}_{k,\text{LoS}}^{\rm MRC} =  \mathrm{SINR}_{k,\text{RF},\text{LoS}}^{\rm MRC} \times \frac{\sum\limits_{k' \neq k}^K p^d_{k'} \frac{|\mathbf{\bar h}^H_k\mathbf{\bar h}_{k'}|^2}{M}\beta_{k'} + \frac{\sigma^2_{\text{RF}}}{\rho_{\text{RF}}}}{\sum\limits_{k' \neq k}^K p^d_{k'} \frac{|\mathbf{\bar h}^H_k\mathbf{\bar h}_{k'}|^2}{M}\beta_{k'} + \frac{\sigma^2}{\rho |\Phi|^2}},
\end{equation}
and 
\begin{equation}
     \mathrm{SINR}_{k,\text{LoS}}^{\rm ZF} =  \mathrm{SINR}_{k,\text{RF},\text{LoS}}^{\rm ZF} \times \frac{ \frac{\sigma^2_{\text{RF}}}{\rho_{\text{RF}}}}{\frac{\sigma^2}{\rho |\Phi|^2}}.
\end{equation}

For the two extreme cases under MRC detection, the rate difference between RAQ–MIMO and RF MIMO can be expressed as
\begin{equation}
\begin{split}
     &\Delta R^{\mathrm{MRC}}_{k,\text{LoS}}
     =\begin{cases}
0, & p^d_k \rightarrow \infty,\\
\frac{T-\tau}{T}\log_2(1+\mathrm{SINR}_{k,\text{RF},\text{LoS}}^{\rm MRC}\frac{\rho|\Phi|^2/\sigma^2}{\rho_{\text{RF}}/\sigma^2_{\text{RF}}}), & p^d_k \rightarrow 0.
\end{cases}
\end{split}
\end{equation}
For the ZF scheme, we have
\begin{equation}
\begin{split}
     &\Delta R^{\mathrm{ZF}}_{k,\text{LoS}} 
    = \begin{cases}
\frac{T-\tau}{T}\log_2(\frac{\rho|\Phi|^2/\sigma^2}{\rho_{\text{RF}}/\sigma^2_{\text{RF}}}) & p^d_k \rightarrow \infty,\\
\frac{T-\tau}{T}\log_2(1+\mathrm{SINR}_{k,\text{RF},\text{LoS}}^{\rm ZF}\frac{\rho|\Phi|^2/\sigma^2}{\rho_{\text{RF}}/\sigma^2_{\text{RF}}}), & p^d_k \rightarrow 0.
\end{cases}
\end{split}
\end{equation}

\begin{remark}[Source of Gain in LoS Channels]\label{Gainlos}
    \emph{Unlike the Rayleigh fading case, the gain of RAQ-MIMO in LoS channels originates solely from the normalized noise background, as the channel-estimation gain vanishes in deterministic conditions.}
\end{remark}

\begin{remark}[Power and Aperture Reduction]\label{powerlos}
    \emph{Similar to the Rayleigh case, the transmit power and antenna aperture required by RAQ-MIMO in LoS channels can be reduced by a factor of \(\frac{\rho|\Phi|^2/\sigma^2}{\rho_{\text{RF}}/\sigma^2_{\text{RF}}}\), thereby allowing proportionally longer communication distances.}
\end{remark}

\subsubsection{Typical Satellite Channel}
Beyond the Rayleigh fading and pure LoS cases discussed above, we now consider a typical satellite scenario, where the received RF power is dominated by LoS propagation, i.e., \(\delta_k \in[10,100] \). Furthermore, given the limited link budget (150-160 dB) and transmit power (23 - 30 dBm), the background noise level is generally larger than \(p^d_k\beta_k,\forall k\). Under these conditions, the typical satellite link can be equivalently modeled as a low-power, LoS-dominated regime. Consequently, the performance gain of RAQ-MIMO over RF MIMO can be approximated as
\begin{equation}
	\frac{T-\tau}{T}\log_2(1+\text{SINR}^{\text{MRC}/\text{ZF}}_{k,\text{RF},\text{LoS}} \frac{\rho |\Phi|^2/\sigma^2}{\rho_{\text{RF}}/\sigma^2_{\text{RF}}}).
\end{equation} 
In satellite channels, the gain of RAQ–MIMO over RF MIMO converges to a constant determined by the normalized noise background \(\frac{\rho|\Phi|^2/\sigma^2}{\rho_{\text{RF}}/\sigma^2_{\text{RF}}}\). The ``squaring'' effect observed under Rayleigh fading vanishes because the channel is nearly deterministic. In other words, for users with weaker LoS components (lower Rician factors), the achievable gain lies between the squaring-law and constant-gain regimes and can even outperform users with stronger LoS components.

\section{Simulation Results}
In this section, Monte Carlo simulations are conducted to validate the derived closed-form expressions and to evaluate the effects of channel estimation accuracy and the Rician factor on system performance. We then highlight the advantages of RAQ-MIMO over conventional RF MIMO and provide a comprehensive performance analysis.

\subsection{Parameter Settings}
We consider the four-level electron transition scheme \(6{\rm S}_{1/2}\rightarrow6{\rm S}_{3/2}\rightarrow47{\rm D}_{5/2}\rightarrow48{\rm P}_{3/2}\). To enhance detection sensitivity, the atomic vapor density is set to \(9 \times 10^{10}\) \(\mathrm{cm}^{-3}\). Unless otherwise specified, the parameters follow those listed in Table I in \cite{gong2024rydberg}. A total of \(K=10\) users are randomly distributed within a circular area of radius 100~m, and their signals are received by a RAQ-MIMO LEO satellite positioned at an altitude of 550~km above the center of the user region. The Rician factors, \(\delta_k, \forall k\), are generated according to Table 6.7.2-1a in \cite{3GPP}. The large-scale fading factors (in dB) is modeled as \(92.45+20\lg(d_k) + 20\lg(f_c)\), where \(d_k\) (in km) is the distance between user \(k\) and the RAQR, and \(f_c\) (in GHz) is the carrier frequency. The small-scale fading factors are drawn from a circularly symmetric complex Gaussian distribution with zero mean and unit variance.

Unless otherwise stated, the user antenna gain is set to \(5\) dBi, and all simulation results are averaged over \(10^{4}\) independent channel realizations.

\subsection{Channel Estimation}
\begin{figure*}
	\centering
	\subfigure[MSE.]{
		\begin{minipage}[t]{0.45\linewidth}
			\centering
			\includegraphics[width=2.75in]{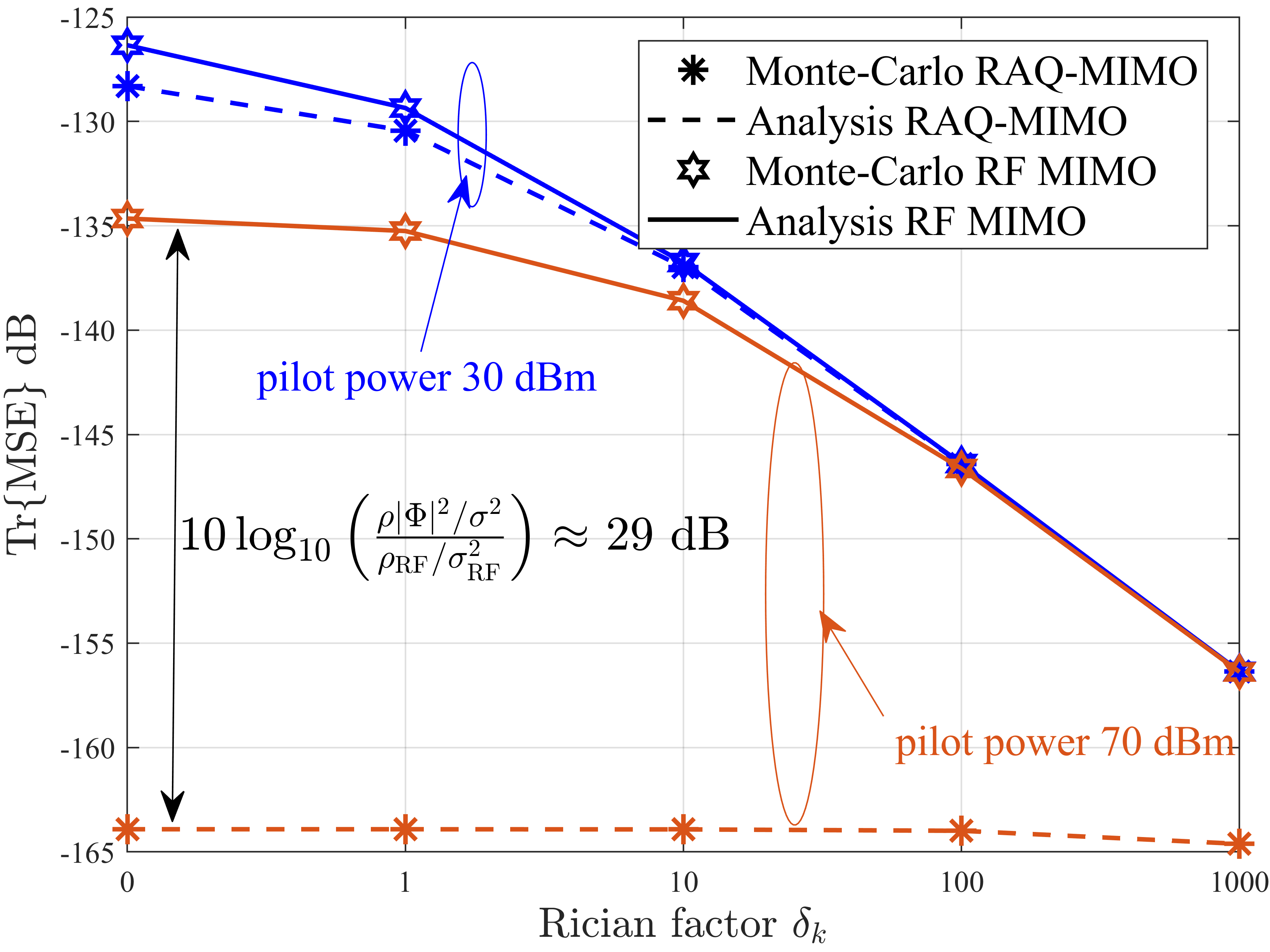}\hspace{10mm}
	\end{minipage}}
	\quad
\subfigure[NMSE .]{\begin{minipage}[t]{0.45\linewidth}
		\centering
		\includegraphics[width=2.75in]{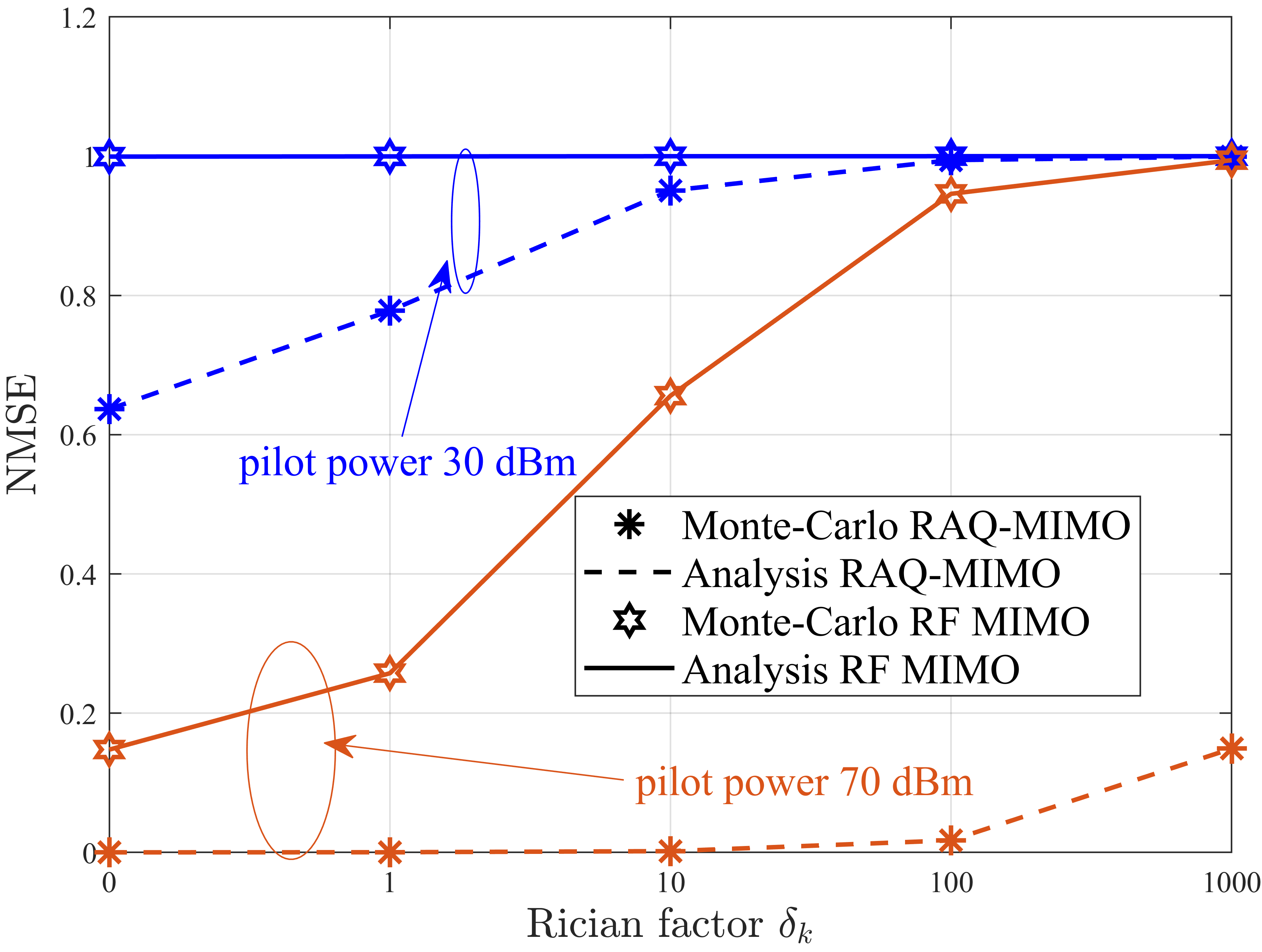}\hspace{10mm}
\end{minipage}}

\caption{Channel estimation under various Rician factors with \(\tau = K\).}
\label{channelestimation}
\end{figure*}

We first evaluate the MSE and NMSE performance of the proposed RAQR with a fixed number of atomic sensors \(M = 900\) and compare the results with those of conventional RF MIMO, as shown in Fig. \ref{channelestimation}. The Monte Carlo simulations closely match the analytical results. As expected, the MSE decreases with increasing Rician factor, while the NMSE exhibits the opposite trend.  This behavior arises because, under strong LoS propagation, background noise dominates the channel variations, making it more challenging to accurately estimate the relative fluctuations even at high power levels. 

Compared with conventional RF MIMO, the RAQ-MIMO demonstrates substantially improved channel estimation performance owing to its lower normalized noise background \(\sigma^2/(\rho|\Phi|^2)\). This confirms that reliable CSI can be effectively obtained even under stringent link-budget constraints. Moreover, in the high-power regime, the performance gain of RAQR over conventional MIMO reaches approximately 29 dB,  consistent with Corollary~1 in Section~II.

\begin{figure}[t]
    \centering
    \includegraphics[width=2.75in]{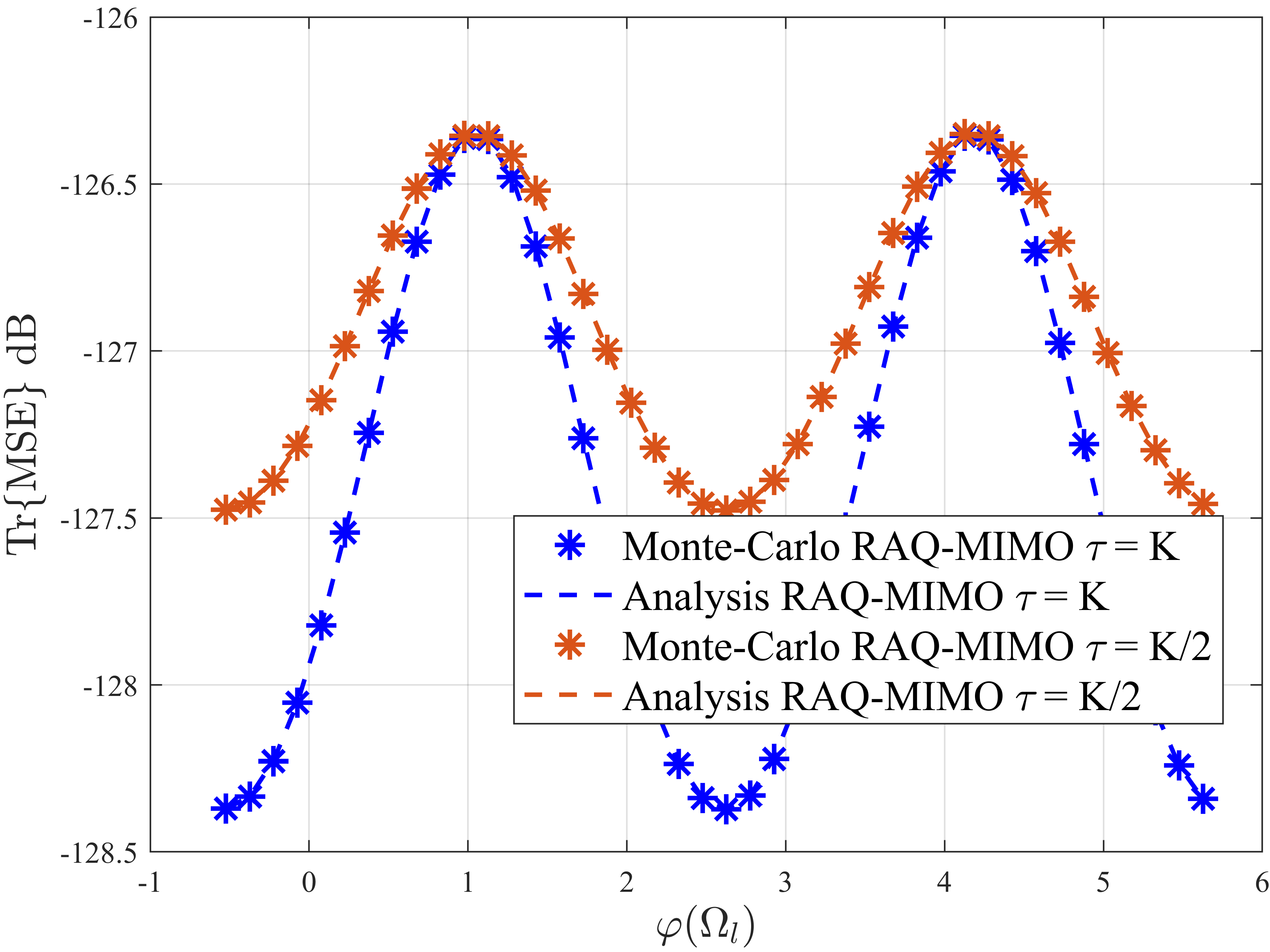}
    \caption{MSE under various \(\varphi(\Omega_l)\).}
    \label{fig:placeholder1}
\end{figure}

Next, with a fixed pilot power of 30~dBm, we investigate the impacts of \(\Phi\) and pilot length \(\tau\) on channel estimation under Rayleigh fading, as shown in Fig. \ref{fig:placeholder1}. The results indicate that the MSE can be effectively minimized through proper configuration of the LO phase, thereby validating Remark~2.

\subsection{Tightness between Analytical Lower Bounds and Monte Carlo Simulations}

\begin{figure}[ht]
	\centering
	\includegraphics[width=2.75in]{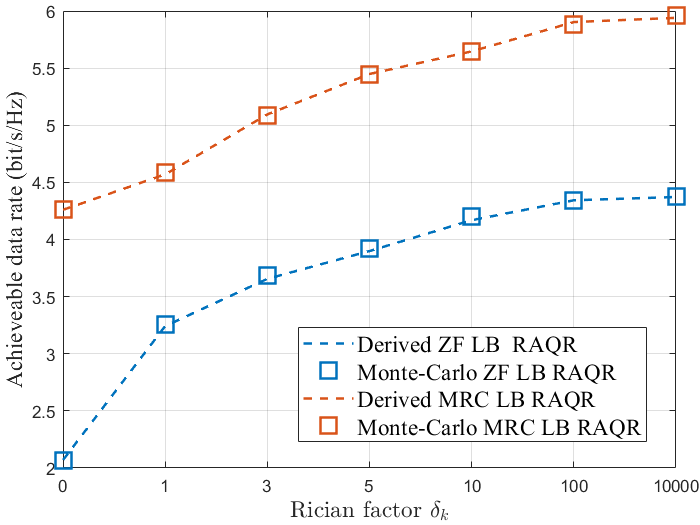}
	\caption{Achievable rate under different Rician factors \( \delta_k, \forall k\).}
	\label{fig:placeholder2}
\end{figure}

As shown in Fig.~\ref{fig:placeholder2}, with the pilot length set to \(\tau = K\) and uniform pilot and data powers \(p^p_k = p^d_k = 30 \) dBm, \(\forall k\), the derived closed-form lower bounds are validated against Monte Carlo simulations.
The analytical results exhibit excellent agreement with the simulation outcomes averaged over \(10^4\) trials, confirming the tightness of the derived expressions. Moreover, the achievable rate increases with the Rician factor, since higher \(\delta_k\) values correspond to more deterministic channels, reduced beamforming mismatch, and improved overall performance.

\subsection{Impacts of the Number of Atomic Sensors}

\begin{figure}[ht]
	\centering
	\includegraphics[width=2.75in]{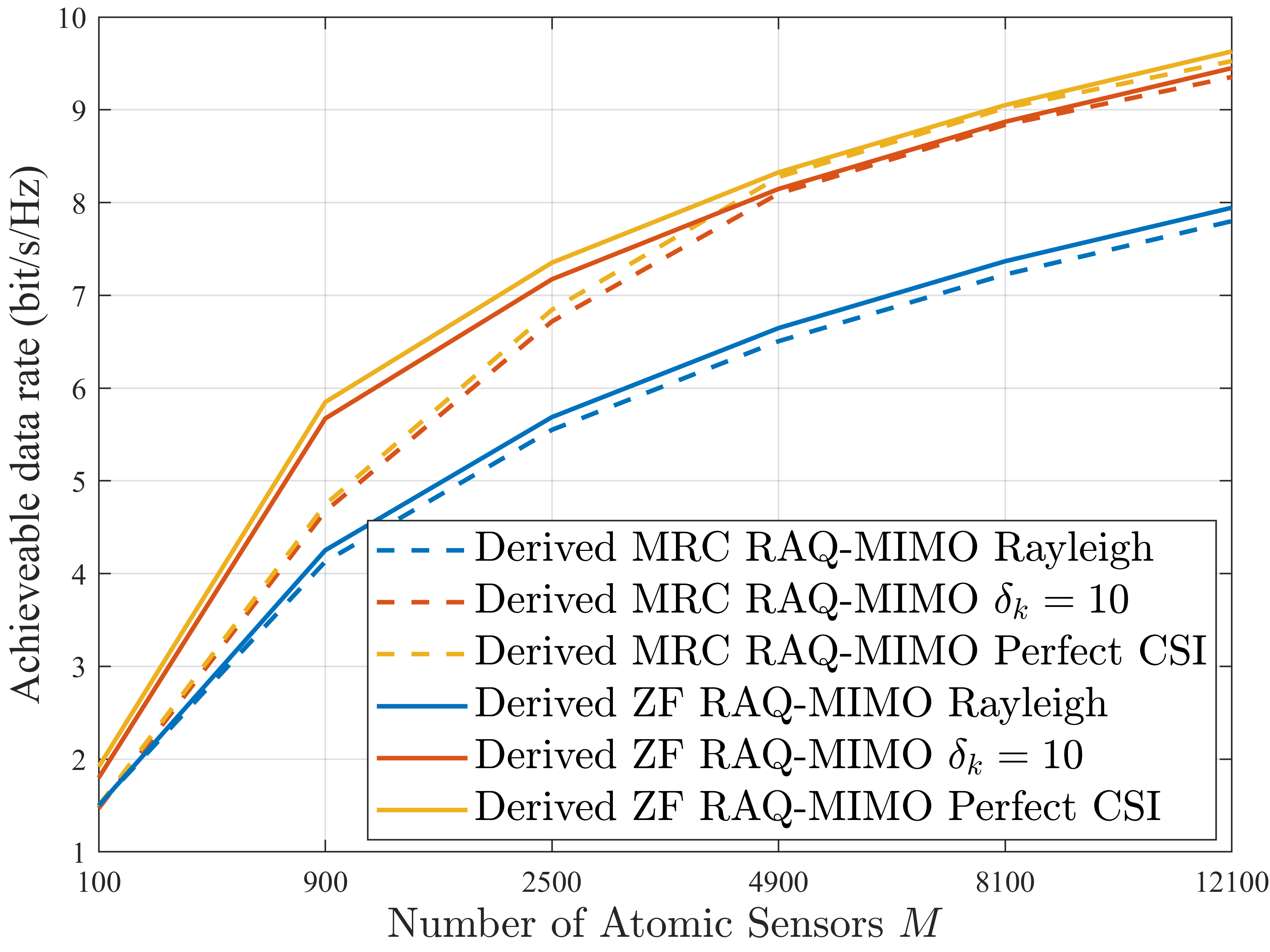}
	\caption{Achievable rate under various atomic sensors \(M\) with \(\tau = K\) and \(p^p_k = p^d_k = 30 \) dBm, \(\forall k\).}
	\label{figimperfectcsi}
\end{figure}

Based on the derived lower bounds, Fig.~\ref{figimperfectcsi} shows the impact of the number of atomic sensors \(M\) on the achievable rate. As \(M\) increases, the rate improves monotonically, reflecting the array-gain benefits analogous to those in conventional RF MIMO systems with larger antenna arrays. It can also be observed that the performance degradation caused by channel-estimation errors is pronounced under Rayleigh fading, whereas in satellite channels the impact of imperfect CSI is marginal due to the dominance of LoS propagation.
\subsection{Performance Comparison}
We now evaluate the benefits of integrating RAQ–MIMO into satellite systems by comparing its performance with that of conventional RF MIMO.

\subsubsection{Non-Zero Rate}
\begin{figure*}[ht]
	\centering
	\subfigure[Rayleigh fading channels.]{
		\begin{minipage}[t]{0.45\linewidth}
			\centering
			\includegraphics[width=2.75in]{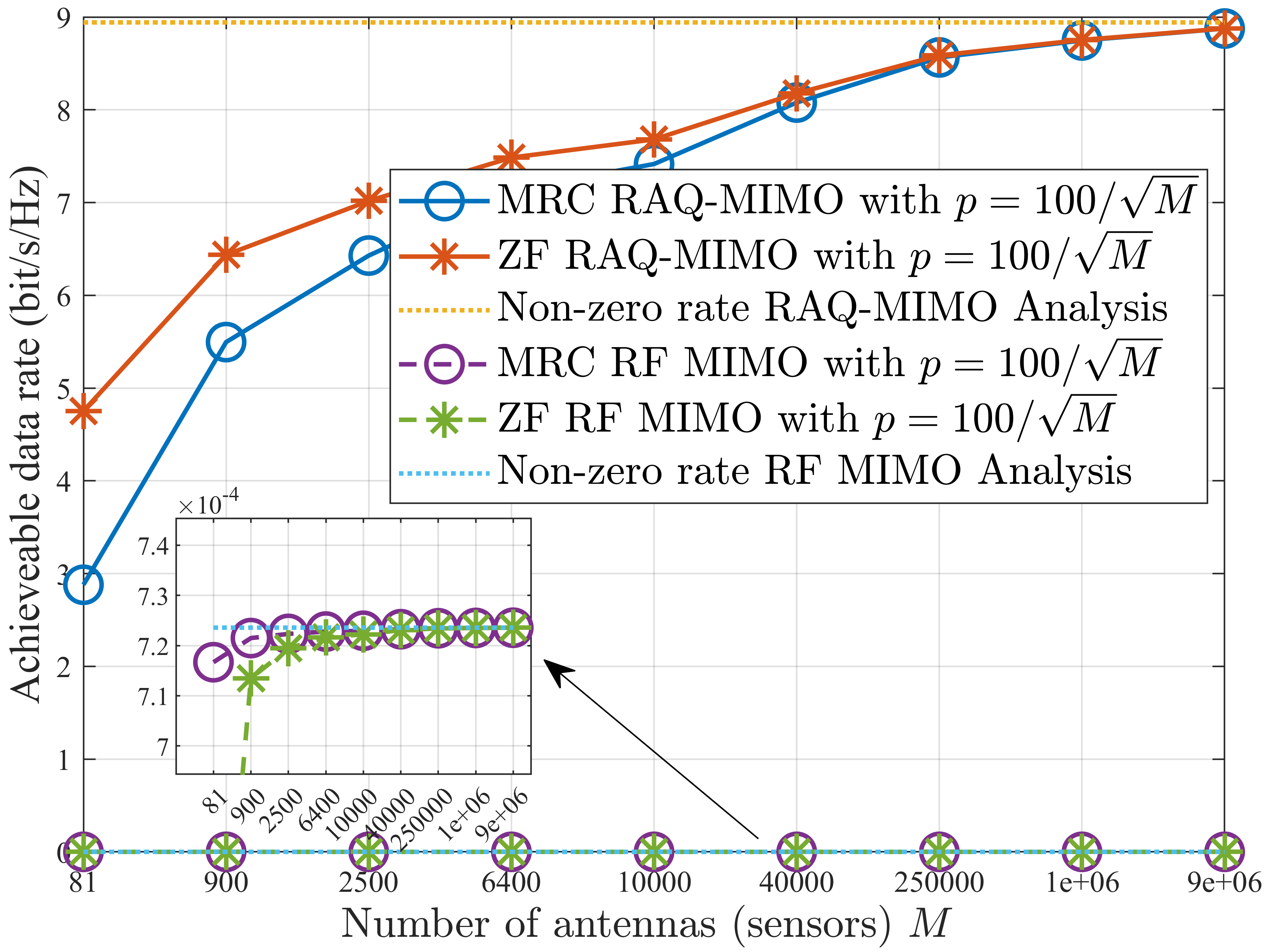}\hspace{10mm}
	\end{minipage}}
	\quad
\subfigure[Typical satellite channels \(\delta_k= 10\), \(\forall k\).]{\begin{minipage}[t]{0.45\linewidth}
		\centering
		\includegraphics[width=2.75in]{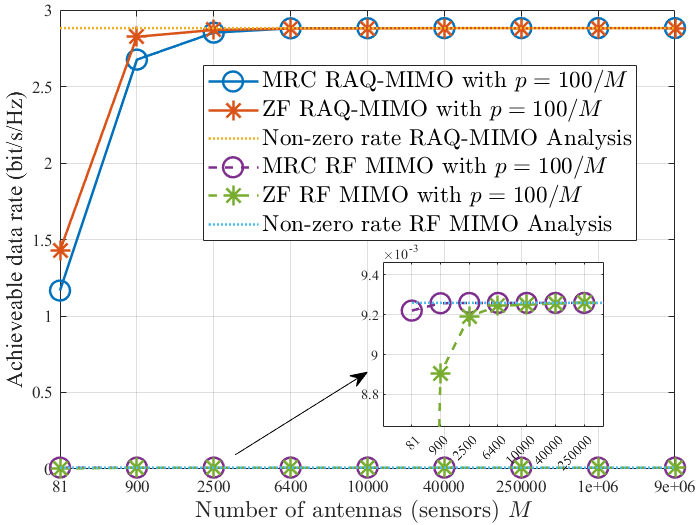}\hspace{10mm}
\end{minipage}}
\caption{Non-zero rate versus the number of antennas (sensors) with \(P = 20\) dBW.}
\label{nonzerorate}
\end{figure*}
Fig.~\ref{nonzerorate} illustrates the power-scaling behavior of RAQR. As observed, a non-zero rate is maintained when the per-user power is scaled as \(p\triangleq p^p_k = p^d_k = \frac{100}{\sqrt{M}}\) under Rayleigh fading and power \(p\triangleq p^d_k = \frac{100}{{M}}\) under typical satellite channels. These results confirm that the classical massive-MIMO power-scaling law remains valid for RAQ–MIMO, consistent with Remark~4. 

Furthermore, the non-varnishing rate under Rayleigh fading exceeds that of the satellite channels, since the gain originates from both improved channel estimation and the reduced normalized noise background. More importantly, the non-vanishing rate achieved by RAQ-MIMO is substantially higher than that of RF MIMO, which indicates that RAQ-MIMO satellites would be a promising solution for ground-to-satellite direct access.

\begin{figure*}[ht]
	\centering
	\subfigure[Rayleigh fading channels.]{
		\begin{minipage}[t]{0.45\linewidth}
			\centering
			\includegraphics[width=2.75in]{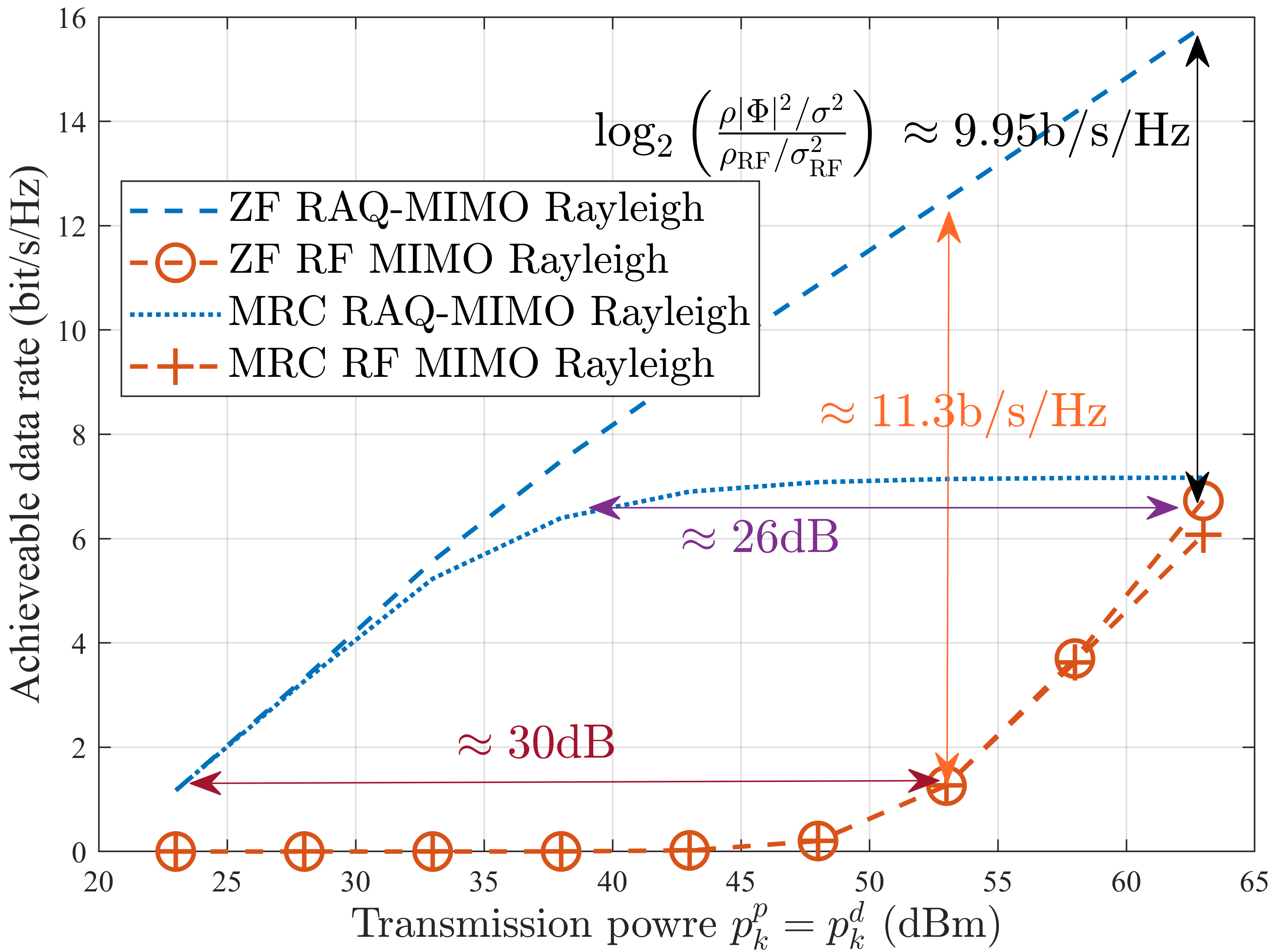}\hspace{10mm}
	\end{minipage}}
	\quad
\subfigure[Typical satellite channels \(\delta_k = 10\), \(\forall k\).]{\begin{minipage}[t]{0.45\linewidth}
		\centering
		\includegraphics[width=2.75in]{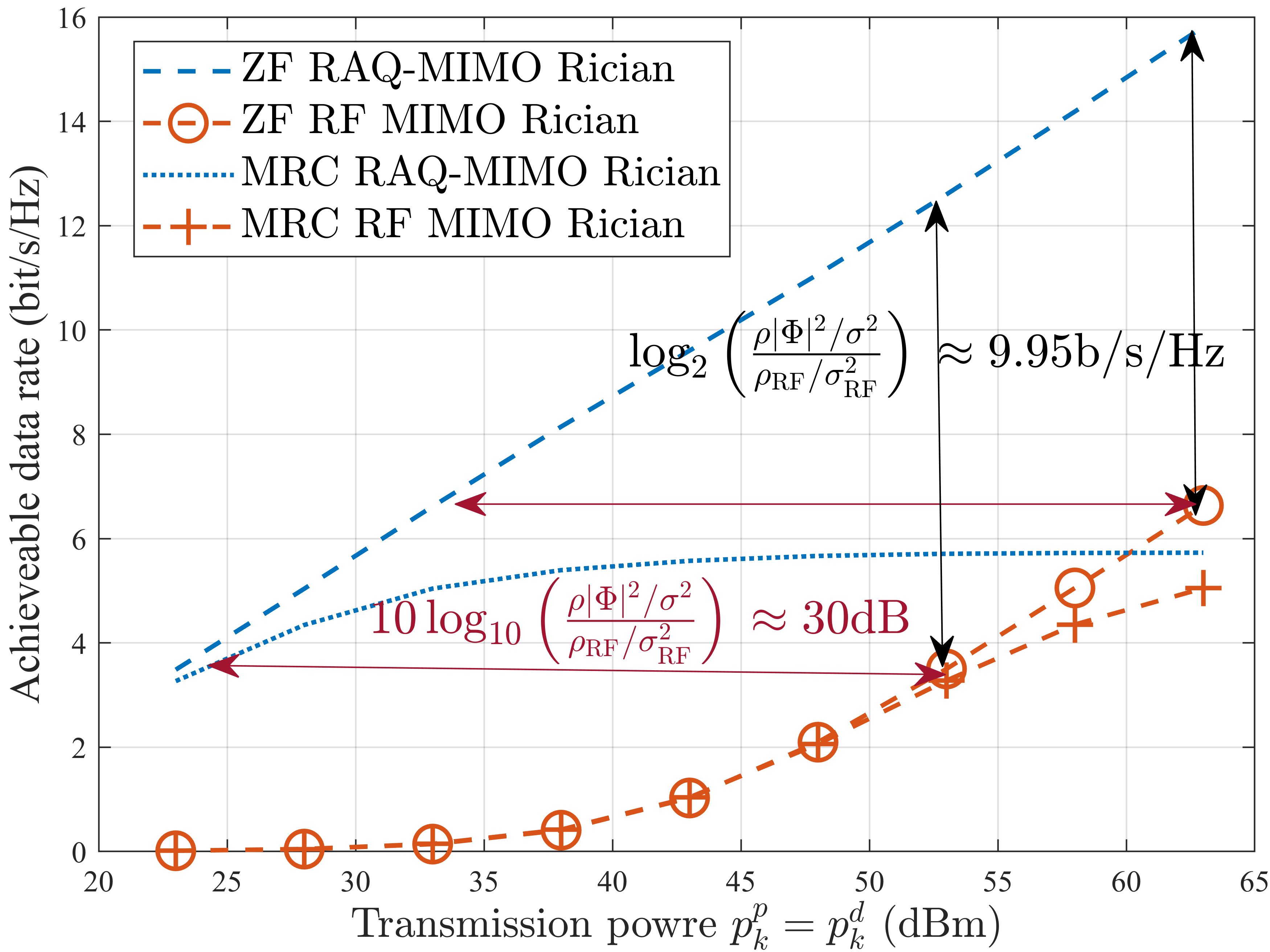}\hspace{10mm}
\end{minipage}}
\caption{Achievable rate under various transmission power  with \(M = 900\).}
\label{fig:transmissionp}
\end{figure*}

\subsubsection{Transmit Power}
Fig.~\ref{fig:transmissionp} quantifies the performance gains of RAQ–MIMO over conventional RF MIMO under both Rayleigh and typical satellite channels. In the high-power regime, the spectral-efficiency gap between RAQ–MIMO and RF MIMO approaches 9.9 bit/s/Hz, consistent with the analytical results in Section~III. At low transmit powers, RAQ–MIMO achieves spectral efficiencies of 11.3 and 9.0 bit/s/Hz under Rayleigh and Rician fading, respectively. The difference stems from additional channel-estimation gains in the Rayleigh case. 

These significant gains confirm that RAQ–MIMO is a promising solution for direct ground-to-satellite uplink access, particularly for handheld or power-limited terminals. Furthermore, for a fixed target rate, the required transmit-power reduction approaches 30~dB in Rician (LoS-dominant) channels, while in Rayleigh fading, the power savings are most pronounced in the low-power region, which verifies our Corollary \ref{Powerray} and Remark \ref{powerlos}. Intuitively, a strong LoS component stabilizes beamforming and channel estimation, maintaining a constant gain determined by the normalized noise background. However, under Rayleigh fading, the benefit of improved channel estimation diminishes as the SNR increases.

\subsubsection{Propagation Distance}
\begin{figure*}[ht]
	\centering
	\subfigure[Rayleigh fading channels.]{
		\begin{minipage}[t]{0.45\linewidth}
			\centering
			\includegraphics[width=2.75in]{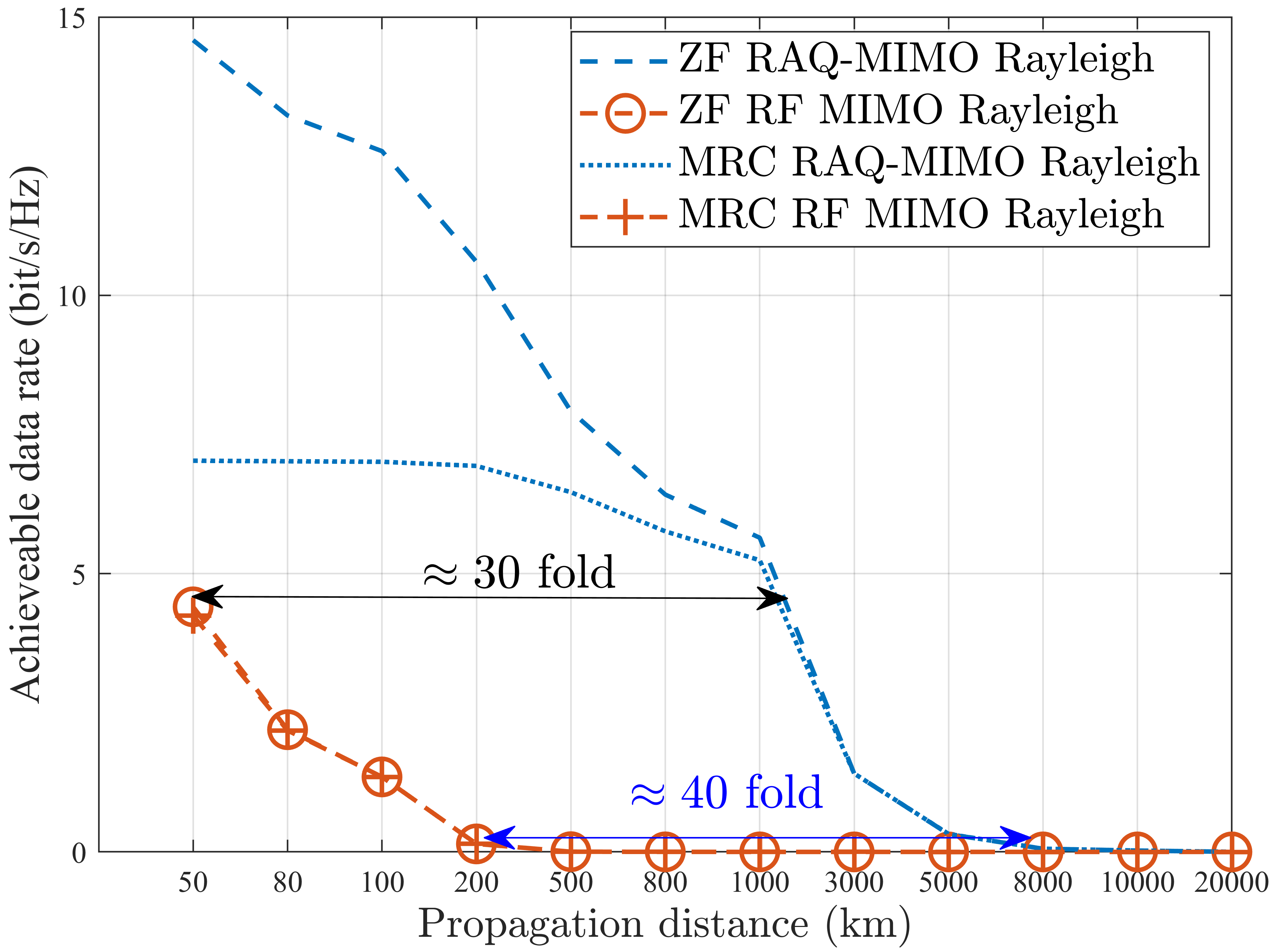}\hspace{10mm}
	\end{minipage}}
	\quad
\subfigure[Typical satellite channels \(\delta_k = 10\), \(\forall k\).]{\begin{minipage}[t]{0.45\linewidth}
		\centering
		\includegraphics[width=2.75in]{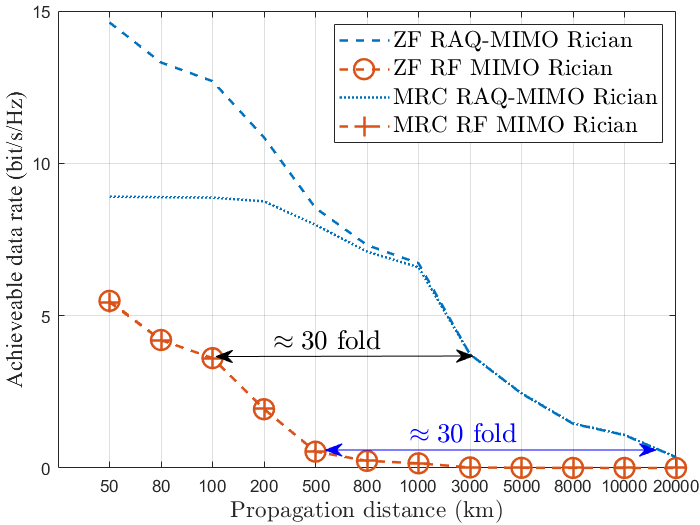}\hspace{10mm}
\end{minipage}}
\caption{Achievable rate versus propagation distance with \(p^p_k = p^d_k = 33\) dBm.}
\label{fig:propagation}
\end{figure*}

By enforcing \(\text{SINR}^{{\rm MRC}/{\rm ZF}}_{k} = \text{SINR}^{{\rm MRC}/{\rm ZF}}_{k,{\rm RF}}\), Fig.~\ref{fig:propagation} illustrates the achievable data rate versus propagation distance for a fair comparison between RAQ–MIMO and RF MIMO. As shown, under Rayleigh fading, the achievable coverage range can be extended by approximately 30-40 times, validating Remark~6. In contrast, for typical satellite (LoS-dominant) channels, the distance extension is around 30-fold, consistent with the inference of Remark~8. These substantial coverage gains demonstrate the superiority of RAQ–MIMO over RF MIMO, thereby positioning RAQ–MIMO as a promising solution for direct ground-to-satellite access and broader space communications.

\subsubsection{Number of Antennas}
\begin{figure*}[ht]
	\centering
	\subfigure[Rayleigh fading channels.]{
		\begin{minipage}[t]{0.45\linewidth}
			\centering
			\includegraphics[width=2.75in]{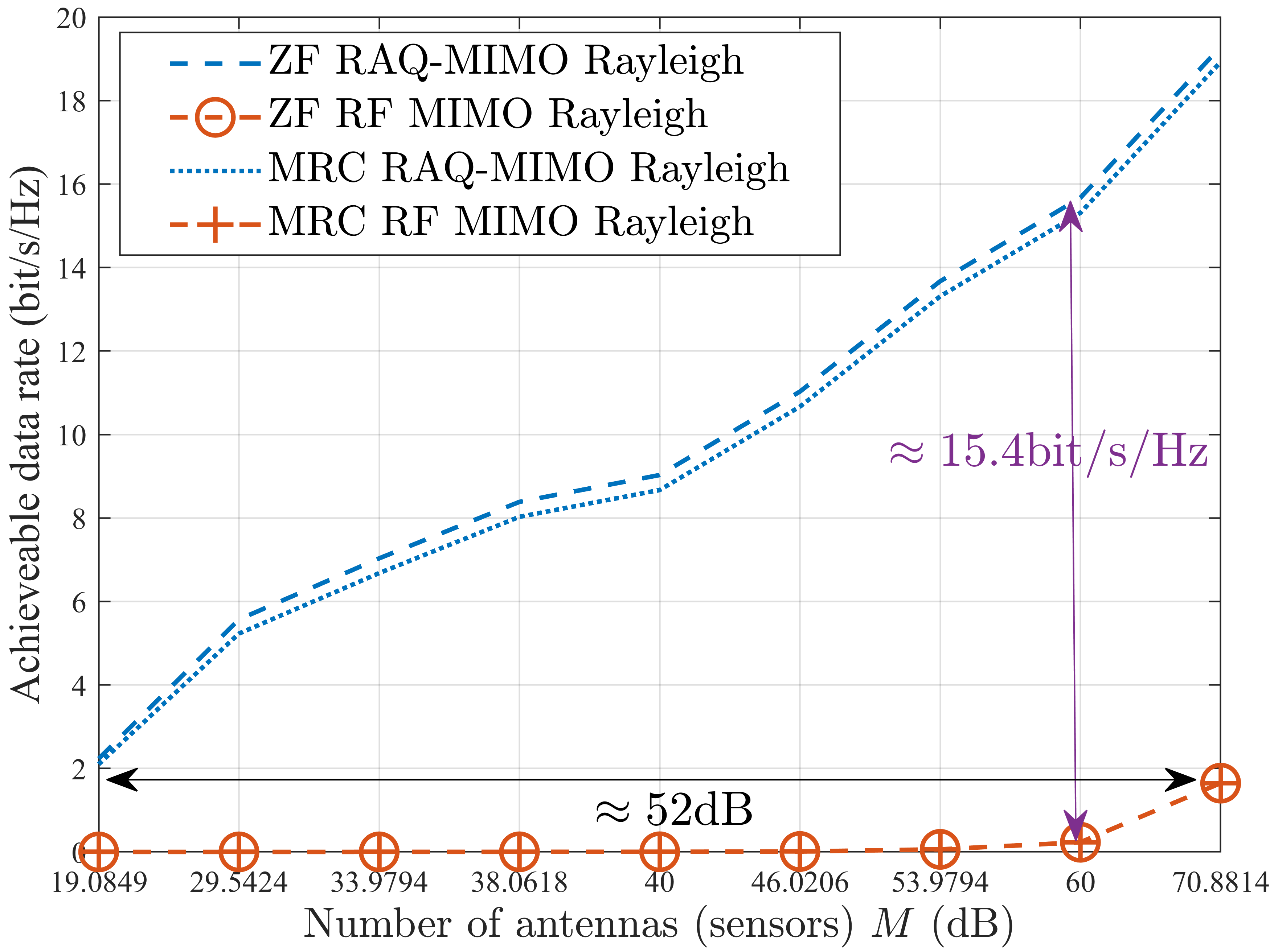}\hspace{10mm}
	\end{minipage}}
	\quad
\subfigure[Typical satellite channels \(\delta_k = 10\), \(\forall k\).]{\begin{minipage}[t]{0.45\linewidth}
		\centering
		\includegraphics[width=2.75in]{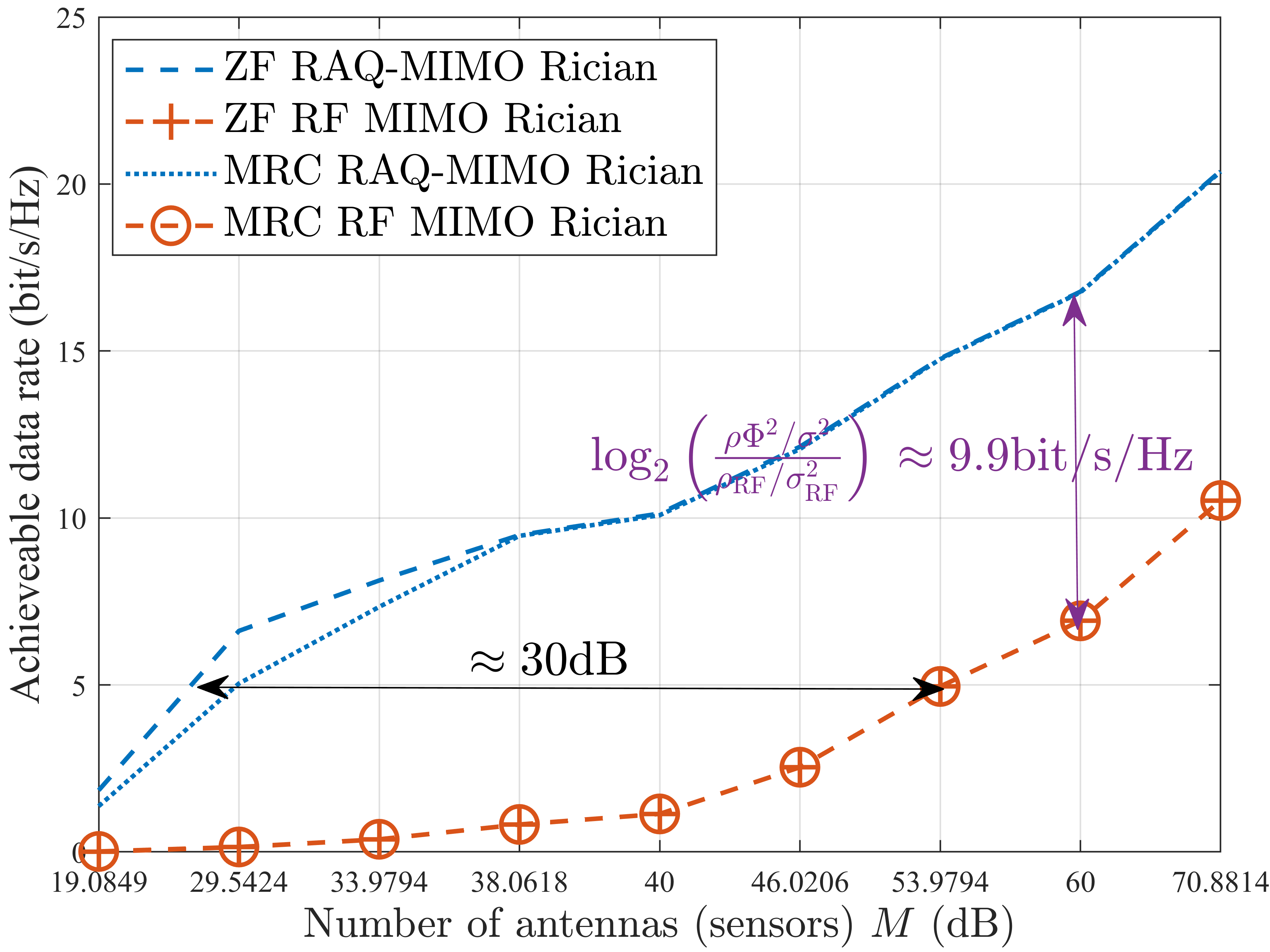}\hspace{10mm}
\end{minipage}}
\caption{Achievable rate under various number of antennas with \(p^p_k = p^d_k = 33\) dBm.}
\label{fig:antennas}
\end{figure*}

Fig.~\ref{fig:antennas} illustrates the achievable rates as a function of the number of antennas. Under Rayleigh fading, the performance gap between RAQ-MIMO and RF MIMO reaches 15.4 bit/s/Hz, which is larger than that observed under Rician fading. This is attributed to additional channel-estimation gains, consistent with the ``squaring effect" described in Remark~6. Owing to this effect, the required number of atomic sensors can be reduced by a factor corresponding to an array-gain difference of approximately 59 dB compared with RF MIMO. 

However, this phenomenon does not persist in the satellite (LoS-dominant) scenario, where channel-estimation gain becomes negligible, corroborating Remarks~7 and~8 in Section~III. Even so, RAQ–MIMO still enables significant reductions in antenna count under LoS-dominant satellite conditions, thereby alleviating mass and volume constraints.



\section{Conclusion}
The paper analyzed the uplink performance and ground-to–satellite direct access enabled by RAQ–MIMO. We first evaluated the channel estimation performance of RAQ-MIMO and demonstrated its superiority over conventional RF MIMO, particularly under typical satellite channels. Based on the estimated channels, closed-form expressions for the lower-bound achievable rates with MRC and ZF detection were derived. Through rigorous analysis, we quantified the performance gap between RAQRs and conventional MIMO systems, confirming the advantages of the proposed RAQ–MIMO architecture.

For typical satellite systems, although the ``squaring” effect characteristic of Rayleigh fading cannot be fully exploited, substantial gains over conventional RF MIMO still persist, primarily due to the improved normalized noise background. Finally, our results revealed that RAQ-MIMO can substantially reduce the required transmit power and antenna aperture, and simultaneously extend communication coverage. Finally, Monte-Carlo simulations validated the analytical derivations and confirmed that RAQ–MIMO can effectively enhance direct ground-to-satellite access.

\begin{appendices}	
\section{Proof of theorem \ref{MRC_SINR_T}}
\label{Prooftheorem1}

First, we define the estimation as 

\begin{equation}
\begin{split}
        \mathbf{\hat h}_k &= \underbrace{\sqrt{\delta_k \alpha_k} \mathbf{\bar h}_k}_{\mathbf{h}_{k,1}} + \underbrace{\frac{\rho \tau p^p_k \alpha^{3/2}_k |\Phi|^2}{\rho \tau p^p_k \alpha_k |\Phi|^2 + \sigma^2}\mathbf{\tilde h}_k}_{\mathbf{h}_{k,2} \triangleq \mu_k \sqrt{\alpha}_k \mathbf{\tilde h}_k} \\
        &+ \underbrace{\frac{\sqrt{\rho \tau p^p_k }\alpha_k \Phi^H}{\rho \tau p^p_k \alpha_k |\Phi|^2 + \sigma^2}\mathbf{D}^H\mathbf{Wq}_k}_{\mathbf{n}_k} .
\end{split}
\end{equation}

Owing to \(\mathbf{c}_k = \Phi\mathbf{D}\mathbf{\hat h}_k\), we have
\begin{equation}
    \small 
    \begin{split}
            &\mathbb{E}\{\sqrt{\rho p^d_k}\Phi\mathbf{c}^{H}_k\mathbf{D}\mathbf{h}_k\} \\
            =&  \mathbb{E}\{\sqrt{\rho p^d_k}|\Phi|^2\mathbf{\hat h}^{H}_k\mathbf{h}_k\} \\
            =& M\sqrt{\rho  p^d_k}|\Phi|^2\Big( \frac{\rho \tau p^p_k\alpha^2_k|\Phi|^2}{\rho\alpha_k \tau p^p_k|\Phi|^2+\sigma^2}  + \delta_k\alpha_k \Big).
    \end{split}
\end{equation}

Next, the power of the leaked signal can be expressed as
\begin{equation}
    \small 
    \begin{split}
        |\text{Ls}_k|^2 &= \mathbb{E}{\{|\sqrt{\rho p^d_k}\Phi\mathbf{c}^{H}_k\mathbf{D}\mathbf{h}_k -\text{Ds}_k|^2\}}  \\
        & = \rho p^d_k |\Phi|^4 \mathbf{h}^H_{k,1}\mathbf{h}_{k,1}\mathbf{h}^H_{k,1}\mathbf{h}_{k,1} + \rho p^d_k|\Phi|^4 \alpha_k |\mathbf{h}^H_{k,1}\mathbf{\tilde h}_{k} |^2 \\
        & + \rho p^d_k |\Phi|^4 \mu^2_k \alpha_k|\mathbf{\tilde h}^H_{k}\mathbf{ h}_{k,1} |^2 + \rho p^d_k |\Phi|^4 \mu^2_k \alpha^2_k|\mathbf{\tilde h}^H_{k}\mathbf{ \tilde h}_{k} |^2 \\
        & + \rho p^d_k |\Phi|^4 |\mathbf{n}^H_k \mathbf{h}_{k,1}|^2 + \rho p^d_k |\Phi|^4 \alpha_k |\mathbf{n}^H_k \mathbf{\tilde h}_{k}|^2 \\
        & + 2\rho p^d_k |\Phi|^4 \mu_k\alpha_k \mathbf{h}^H_{k,1} \mathbf{h}_{k,1}\mathbf{\tilde h}^H_{k}\mathbf{\tilde h}_{k} - |\text{Ds}_k|^2\\
       & = \rho p^d_k |\Phi|^4 \Big(M^2 \delta^2_k \alpha^2_k+ M\delta_k \alpha^2_k +M\mu^2_k \delta_k \alpha^2_k \\
       & + (M^2+ M)\mu^2_k \alpha^2_k+M\delta_k\alpha_k\frac{\mu^2_k}{\rho \tau p^d_k |\Phi|^2} \sigma^2 \\&+ M \alpha_k\frac{\mu^2_k}{\rho \tau p^d_k |\Phi|^2} \sigma^2 
       + 2\mu_k\delta_k \alpha^2_kM^2  \Big )  \\&- M^2\rho p^d_k |\Phi|^4(\mu_k\alpha_k + \delta_k \alpha_k)^2 \\
       & = M \rho p^d_k |\Phi|^4 \Big( \delta_k \alpha_k^2 +\delta_k \mu_k\alpha^2_k + \mu_k\alpha_k^2 \Big) \\
       & = M \rho p^d_k |\Phi|^4 \Big( \beta_k\alpha_k\mu_k + \delta_k \alpha_k^2   \Big)
    \end{split}  
\end{equation}

The inter-user interference is given by
\begin{equation}
    \small 
    \begin{split}
        |\text{UI}_{k,k'}|^2 &= \mathbb{E}{\{|\sqrt{\rho p^d_{k'}}\Phi\mathbf{c}^{H}_k\mathbf{D}\mathbf{h}_{k'}|^2\}}\\
        & = \rho p^d_{k'} |\Phi|^4 \mathbb{E}\{\mathbf{\hat h}^H_{k}\mathbf{h}_{k'}\mathbf{h}^H_{k'} \mathbf{\hat h}_k\}\\
        & = \rho p^d_{k'} |\Phi|^4 \Big(\mathbf{h}^H_{k,1}\mathbf{h}_{k',1}\mathbf{h}^H_{k',1}\mathbf{h}_{k,1}+\alpha_{k'}\mathbf{h}^H_{k,1}\mathbf{\tilde h}_{k'}\mathbf{\tilde  h}^H_{k'}\mathbf{h}_{k,1}\\
        & + \mu_k^2\alpha_k\mathbf{\tilde h}^H_k\mathbf{h}_{k',1}\mathbf{h}^H_{k',1}\mathbf{\tilde h}_k+\mu_k^2\alpha_k\alpha_{k'}  \mathbf{\tilde h}^H_k \mathbf{\tilde h}_{k'}\mathbf{\tilde h}^H_{k'}\mathbf{\tilde h}_k \\
        & + \mathbf{n}^H_k\mathbf{h}_{k',1}\mathbf{h}^H_{k',1}\mathbf{n}_k + \alpha_{k'}\mathbf{n}^H_k\mathbf{\tilde h}_{k'}\mathbf{\tilde h}^H_{k'}\mathbf{n}_k\Big)\\
        & = \rho p^d_{k'} |\Phi|^4(\mathbf{h}^H_{k,1}\mathbf{h}_{k',1}\mathbf{h}^H_{k',1}\mathbf{h}_{k,1}+M \alpha_{k'}\alpha_k\delta_k+M \beta_{k'}\alpha_k\mu_k).
    \end{split}  
\end{equation}

Similarly, the noise can be written as
\begin{equation}
    \small 
    \begin{split}
        |\text{N}_{k}|^2 &= \mathbb{E}{\{|\mathbf{c}^{H}_k\mathbf{w}|^2\}}\\
        & = M|\Phi|^2\sigma^2 (\mu_k\alpha_k + \delta_k\alpha_k).
    \end{split}  
\end{equation}

Finally, we complete this proof by substituting the above results into (\ref{MRC_LB_rate}).

\section{Proof of theorem \ref{ZF_SINR_T}}
\label{Prooftheorem2}

To derive the lower bound, the term \(\mathbb{E}\{[\mathbf{\hat H}^H\mathbf{\hat H}]_{k,k}\}\) needs to be tackled, where \(\mathbf{\hat H} = [\mathbf{\hat h}_1,\cdots,\mathbf{\hat h}_K]\) collects the estimated channels and \([\mathbf{X}]_{k,k}\) denotes the \((k,k)\)-th element of \(\mathbf{X}\). 

First, we prove that \(\mathbf{\hat H}\) follows a Gaussian distribution of \(\mathbf{\hat H} \sim \mathcal{CN}(\mathbf{\bar H},\mathbf{I}_M \otimes \text{Diag}\{{\mu_1\alpha_1},\cdots, {\mu_K\alpha_K}\})\), where \(\mathbf{\bar H} = [\sqrt{\delta_1\alpha_1}\mathbf{\bar h}_1,\cdots,\sqrt{\delta_K\alpha_K}\mathbf{\bar h}_K]\). Then, by using the Lemma 6 in \cite{Kangda2022JSAC}, the product \(\mathbf{\hat H}^H\mathbf{\hat H}\), follows a complex non-central Wishart distribution, which can be given by
\begin{equation}
\label{CW}
    \mathbf{\hat H}^H\mathbf{\hat H} \sim \mathcal{CW}_K(M,\boldsymbol{\Psi},\boldsymbol{\Sigma}),
\end{equation}
where \(\boldsymbol{\Psi} = \text{Diag}\{\mu_1\alpha_1,\cdots,\mu_K\alpha_K\}\) and \(\boldsymbol{\Sigma} = \boldsymbol{\Psi}^{-1}\mathbf{\bar H}^H\mathbf{\bar H} \). Even though the statistical property of (\ref{CW}) is very complex, we approximate the non-central Wishart distribution by a central Wishart distribution \cite{steyn1972approximations,siriteanu2012mimo}. Therefore, the non-central Wishart distribution (\ref{CW}) is approximated by a central one with the same first-order moment, which can be expressed as
\begin{equation}
\label{CW1}
\begin{split}
        \mathbb{E}\{\mathbf{\hat H}^H\mathbf{\hat H}\} &= M\text{Diag}\{\mu_1\alpha_1,\cdots,\mu_K\alpha_K\}+\mathbf{\bar H}^H\mathbf{\bar H}. \\
\end{split}
\end{equation}
The central Wishart distribution with the same mean is given by 
\begin{equation}
    \label{CW2}
    \mathbf{\hat H}^H\mathbf{\hat H} \sim \mathcal{CW}_K(M,\boldsymbol{\Psi}+\frac{\mathbf{\bar H}^H\mathbf{\bar H}}{M}).
\end{equation}
Therefore, we have
\begin{equation}
    \label{CW3}
    \mathbb{E}\Big\{\big(\mathbf{\hat H}^H\mathbf{\hat H} \big)^{-1}\Big\} = \frac{(\boldsymbol{\Psi}+\frac{\mathbf{\bar H}^H\mathbf{\bar H}}{M})^{-1}}{M-K}.
\end{equation}

Owing to \(\mathbf{C} =\Phi\mathbf{D}\mathbf{\hat H} [(\Phi\mathbf{D}\mathbf{\hat H})^H\Phi\mathbf{D}\mathbf{\hat H}]^{-1}\), we have
\begin{equation}
    \small 
    \mathbb{E}\{\sqrt{\rho p^d_k}\Phi\mathbf{c}^{H}_k\mathbf{D}\mathbf{h}_k\} = \sqrt{\rho p^d_k}.
\end{equation}

Then, the leaked signal based on ZF can be expressed as
\begin{equation}
    \small 
    \begin{split}
        |\text{Ls}_k|^2 &= \mathbb{E}{\{|\sqrt{\rho p^d_k}\Phi\mathbf{c}^{H}_k\mathbf{D}\mathbf{h}_k|^2\}} - |\text{Ds}_k|^2 \\
        & = \rho p^d_k  \mathbb{E}\{\mathbf{c}^H_k\mathbf{h}_k\mathbf{h}^H_k \mathbf{c}_k\}-|\text{Ds}_k|^2 \\
        & = \rho p^d_k (\alpha_k - \mu_k\alpha_k)\times \frac{\big[(\boldsymbol{\Psi}+\frac{\mathbf{\bar H}^H\mathbf{\bar H}}{M})^{-1}\big]_{k,k}}{M-K}. 
    \end{split}  
\end{equation}

Next, the inter-user interference and noise can be given by
\begin{equation}
    \small 
    \begin{split}
        |\text{UI}_{k,k'}|^2 &= \mathbb{E}{\{|\sqrt{\rho p^d_{k'}}\Phi\mathbf{c}^{H}_k\mathbf{D}\mathbf{h}_{k'}|^2\}}\\
        & = \rho p^d_{k'} \mathbb{E}\{\mathbf{c}^H_{k}\mathbf{h}_{k'}\mathbf{h}^H_{k'} \mathbf{c}_k\}\\
        & = \rho p^d_{k'} (\alpha_{k'} - \mu_{k'}\alpha_{k'})\times \frac{\big[(\boldsymbol{\Psi}+\frac{\mathbf{\bar H}^H\mathbf{\bar H}}{M})^{-1}\big]_{k,k}}{M-K},
    \end{split}  
\end{equation}
and
\begin{equation}
    \small 
    \begin{split}
        |\text{N}_{k}|^2 &= \mathbb{E}{\{|\mathbf{c}^{H}_k\mathbf{w}|^2\}}\\
        & = \frac{\sigma^2}{|\Phi|^2}\times \frac{\big[(\boldsymbol{\Psi}+\frac{\mathbf{\bar H}^H\mathbf{\bar H}}{M})^{-1}\big]_{k,k}}{M-K}.
    \end{split}  
\end{equation}

By combining the results, the proof of Theorem \ref{ZF_SINR_T} is completed.
\end{appendices}	


\bibliographystyle{IEEEtran}
\bibliography{myref}

\end{document}